\documentclass[10pt,preprint2]{aastex}
\usepackage{amssymb,graphicx,longtable,lscape,natbib,multirow}
\usepackage{wasysym}

\def\msol{\hbox{$\rm\thinspace M_{\odot}\thinspace$}} 
\def\sol{\hbox{$\rm\thinspace _{\odot}\thinspace$}} 
\def\etal{{\it et al.\thinspace}}
\def\eg{{\it e.g.\ }}

\newcommand{\be}{\begin{equation}}
\newcommand{\ba}{\begin{eqnarray}}
\newcommand{\ee}{\end{equation}}
\newcommand{\ea}{\end{eqnarray}}

\newcommand{\nuclei}[2]{\ensuremath{\mathrm{^{#1}#2}}}

\newcommand{\helium}[1][4]{\nuclei{#1}{He}}

\newcommand{\carbon}[1][12]{\nuclei{#1}{C}}

\newcommand{\oxygen}[1][16]{\nuclei{#1}{O}}

\newcommand{\silicon}[1][28]{\nuclei{#1}{Si}}

\newcommand{\nickel}[1][58]{\nuclei{#1}{Ni}}

\begin{document}

\title{\nickel[56] Production in Double Degenerate White Dwarf Collisions}
\author{Cody Raskin\altaffilmark{1}, Evan Scannapieco\altaffilmark{1}, Gabriel Rockefeller\altaffilmark{2}, Chris Fryer\altaffilmark{2}, Steven Diehl\altaffilmark{2}, \& F.X. Timmes\altaffilmark{1,3}}
\altaffiltext{1}{School of Earth and Space Exploration,  Arizona State University, P.O.  Box 871404, Tempe, AZ, 85287-1404}  
\altaffiltext{2}{Los Alamos National Laboratories, Los Alamos, NM 87545}
\altaffiltext{3}{The Joint Institute for Nuclear Astrophysics}

\begin{abstract}

We present a comprehensive study of white dwarf collisions as an avenue for creating type Ia supernovae. Using a smooth particle hydrodynamics code with a 13-isotope, $\alpha$-chain nuclear network, we examine the resulting \nickel[56] yield as a function of total mass, mass ratio, and impact parameter. We show that several combinations of white dwarf masses and impact parameters are able to produce sufficient quantities of \nickel[56] to be observable at cosmological distances. We find the \nickel[56] production in double-degenerate white dwarf collisions ranges from sub-luminous to the super-luminous, depending on the parameters of the collision.  For all mass pairs, collisions with small impact parameters have the highest likelihood of detonating, but  \nickel[56] production is insensitive to this parameter in high-mass combinations, which significantly increases their likelihood of detection. We also find that the \nickel[56] dependence on total mass and mass ratio is not linear, with larger mass primaries producing disproportionately more \nickel[56] than their lower mass secondary counterparts, and symmetric pairs of masses producing more \nickel[56] than asymmetric pairs.

\end{abstract}

\section{Introduction}

While the preferred mechanism for type Ia supernovae (SNeIa) involves a single white dwarf star accreting material from a non-degenerate companion (Whelan \& Iben 1973; Nomoto 1982; Hillebrandt \& Niemeyer 2000), recent observational evidence suggests a non-negligible fraction of observed SNeIa may derive from double-degenerate progenitor scenarios. Scalzo \etal (2010) observed the supernova SN~2007if photometrically, and assuming no host galaxy extinction, they found 1.6$\pm$0.1\msol of \nickel[56] with 0.3-0.5\msol of unburned carbon and oxygen forming an envelope. This \nickel[56] yield implies a progenitor mass of 2.4$\pm$0.2\msol, which is well above the Chandrasekhar limit - the maximum mass for a non-rotating white dwarf (Chandrasekhar 1931, Pfannes \etal 2010, Yoon \& Langer 2004, Yoon \& Langer 2005). It follows that two white dwarfs must have been involved in the event that produced SN 2007if, since a single white dwarf cannot accrete enough material to reach this mass without either exploding as a SNIa or collapsing to form a neutron star (Yoon \etal 2007). Furthermore, spectroscopic observations by Tanaka \etal (2010) suggest SN~2009dc produced $\apprge1.2$\msol of \nickel[56], depending on the assumed dust absorption. This also implies a progenitor mass $>1.4$\msol as 0.92\msol of \nickel[56] is the greatest yield a Chandrasekhar mass can produce (Khokhlov \etal 1993). 

Howell \etal (2006) inferred from their observations of SN~2003fg that $\sim1.3$\msol of \nickel[56] was produced, as did Hicken \etal (2007) in their observations of SN~2006gz. There is a growing body of evidence supporting double-degenerate SNeIa progenitor systems. Since any supernova arising from a double-degenerate progenitor scenario may not fit the standard templates for SNeIa, these transients must be filtered out if SNeIa are to remain as premier cosmological tools. To that end, we must develop models that give clear and detectable signatures of double-degenerate SNeIa to distinguish them from standard SNeIa. 

Currently, models of double-degenerate progenitors are split into two, dynamically different scenarios. In the first, white dwarfs in close binaries lose angular momenta through gravitational radiation, ultimately merging into a thermally supported super-Chandrasekhar object or detonating outright (Iben \& Tutukov 1984; Webbink 1984; Benz \etal 1989a; Yoon \etal 2007; Pakmor \etal 2010). In the second, two white dwarfs collide in dense stellar systems such as globular cluster cores (Raskin \etal 2009; Rosswog \etal 2009; Lor始-Aguilar \etal 2009; Lor始-Aguilar \etal 2010), where white dwarf number densities can be as high as $\approx10^{4}$ pc$^{-3}$. This follows from conservative estimates for the average globular cluster mass, $10^6$\msol (Brodie \& Strader 2006), and for the average globular cluster core radius, 1.5 pc (Peterson \& King 1975), taken together with the Salpeter IMF (Salpeter 1955). Assuming cluster velocity dispersions on the order of 10 km s$^{-1}$, this allows for $10-100$, $z \lesssim 1$ collisions per year. Observations by Chomiuk \etal (2008) of globular clusters in the nearby S0 galaxy NGC~7457 have detected what is likely to be a SNIa remnant. Given the difficulty in distinguishing SNeIa as residing in galaxy field stars or in globular clusters in front of or behind their host galaxies (Pfahl \etal 2009), the frequency with which these can occur warrants investigation. 

Numerical simulations of white dwarf collisions were pioneered in Benz \etal (1989b) using a smooth particle hydrodynamics (SPH) code. They concluded from their results that white dwarf collisions were of little interest as the \nickel[56] yields were small. However, their simulations employed an approximate equation of state for white dwarfs and resolutions were low, relative to what is possible with current computing resources. Moreover, as will be discussed, the infall velocities and velocity gradients play a crucial role in the final \nickel[56] yields. 

More recently, Raskin \etal (2009), Rosswog \etal (2009), and Lor始-Aguilar \etal (2010) revisited collisions using up-to-date SPH codes and vastly more particles ($8\times 10^5$, $2\times 10^6$, and $4\times 10^5$, respectively). In Raskin \etal (2009), a single mass pair (0.6\msol$\times 2$) was explored with three impact parameters, whereas in Rosswog \etal (2009), several mass pairs were examined in direct, head-on collisions. Both of these papers aimed at establishing double-degenerate collisions as SNeIa progenitors, finding that \nickel[56] is indeed produced prodigiously in such collisions, lending credence to their candidacy. Lor始-Aguilar \etal (2010) examined one mass pair (0.6\msol + 0.8\msol), but at a number of different impact parameters, ranging from those that resulted in direct collisions to those that resulted in eccentric binaries, aimed at establishing the parameters of white dwarf coalescence arising from collisional dynamics. 

In this paper, we revisit the three impact parameters studied in Raskin \etal (2009) using a variety of mass pairings. Using 22 combinations of masses and impact parameters, we aim to answer five key questions; how does \nickel[56] production depend on
\begin{itemize}
\item the total mass of the system?
\item the mass ratio of the two stars?
\item the impact parameter?
\item the infall velocities of the constituent stars?
\item tidal effects?
\end{itemize}
While the last two of these questions can be eliminated with robust initial conditions, they are nevertheless important details that are sometimes overlooked. Armed with this information, we will be able to make some conclusions about the observability of different combinations of collision parameters, and to determine whether the resulting transient of any particular collision is as luminous as a SNIa. 

The structure of this paper is as follows. In \S2, we discuss the details of our initial conditions and our new hybrid burning nuclear network. In \S3, we give the details of the results of each simulation that resulted in a detonation along with a study of the effect of numerical parameters on the \nickel[56] yield in \S3.1.2, and in \S4, we discuss those that resulted in remnants. Finally, in \S5, we summarize our results and conclusions. 

\section{Method}

\subsection{Particle Setups \& Initial Conditions}

As in Raskin \etal (2009), we employ a version of a 3D SPH code called SNSPH (Fryer \etal 2006). SPH codes are particularly well suited to these kinds of simulations as the white dwarf stars involved are very dense and moving very rapidly. Advecting rapidly moving, isothermal, cold white dwarfs in Eulerian, grid-based codes introduces perturbations that can be challenging to overcome. Moreover, because many of our simulations are grazing impacts, conservation of angular momentum is crucial to the final outcomes, for which SNSPH excels (Fryer \etal 2006).

In our previous work, we used a Weighted Voronoi Tessellations method (WVT, Diehl \& Statler 2006) for our particle setups. This method arranges particles in a pseudo-random spatial distribution with thermodynamic quantities that are consistent with the chosen equation of state (EOS). The default operation for this method is to allow the masses of particles to vary in order to keep their sizes, or smoothing lengths ($h$), constant throughout the initial setup. This approach has its advantages when it comes to spatial arrangement, but one disadvantage is that it produces a uniform level of refinement in the initial conditions regardless of where most of the mass resides. The result is that much higher particle counts are required to reach convergence. 

To remedy this, we modified the WVT method to keep mass fixed, varying $h$ consistent with the density profiles of white dwarf stars. This has the effect of concentrating resolution where most of the mass resides, vastly reducing the required particle counts for convergence. In fact, whereas in Raskin \etal (2009), we showed that convergence of the \nickel[56] yield was reached at approximately $10^6$ particles, using constant mass particles, we reach convergence with only 200,000. A convergence test on particle count of our fiducial case, 0.64\msol$\times2$ with zero impact parameter, is discussed in \S3.1.2. 

\begin{figure}[ht]
\centering
\includegraphics[width=7cm,height=4cm]{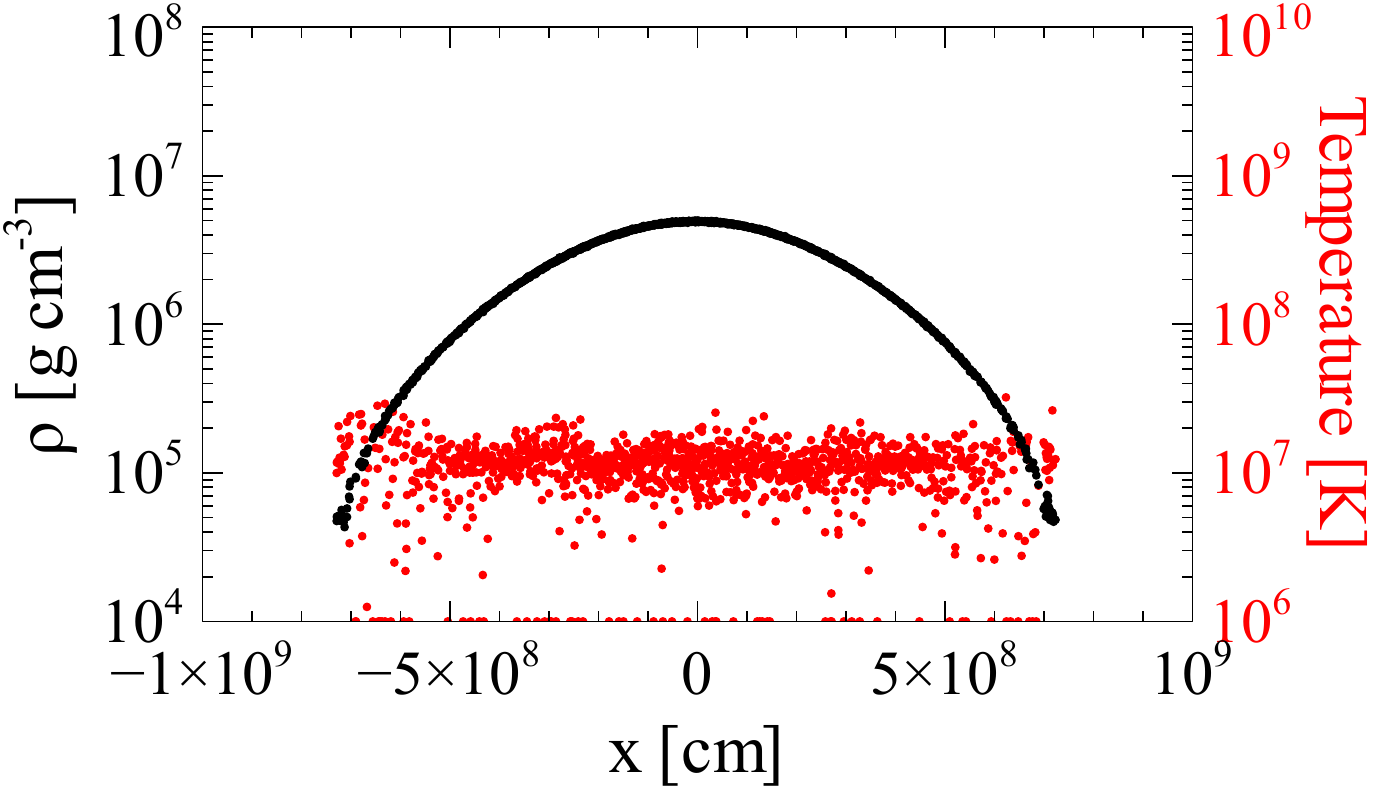}
\caption{Temperatures and densities of particles lying on the x-axis in a 0.64\msol white dwarf. This star was created using WVT and isothermalized to $10^7$K after $\sim 5$ minutes. }
\label{fig:profile}
\end{figure}

A further modification we have added to our previous approach is an isothermalization step in our initial conditions. When mapping 1D profiles for cold white dwarfs onto a resolution limited, 3D particle setup, there is often a relaxation time, during which the stars oscillate before finding equilibrium. For white dwarf masses of $\approx 0.6$\msol, this settling time is short, but for larger masses, the oscillations can continue for several minutes or hours. These repeated gravitational contractions heat the interiors of the stars until they can no longer be considered ``cold" white dwarfs. Therefore, we relaxed each individual star in a modified version of SNSPH that artificially cools the stars by keeping each particle at a constant temperature during the stars' oscillations until they reach a cold equilibrium. Figure \ref{fig:profile} shows a temperature profile for a 0.64\msol white dwarf that has been passed through this isothermalization routine, indicating an isothermal temperature of $\approx 10^7$K throughout.

As in our previous work, the initial conditions for the positions and velocities of the white dwarf stars in our simulations were generated using a fourth-order Runge-Kutta solver with an adaptive time-step that integrates simple kinematic equations. The impactor star was initially given a small velocity comparable to the velocity dispersion of globular cluster cores, $\sigma=10$km/s. The solver places the stars at 0.1R\sol apart with the proper velocity vectors expected for free-fall from large initial separations with a given velocity dispersion. 

\begin{figure}[ht]
\centering
\includegraphics[width=7cm,height=4cm]{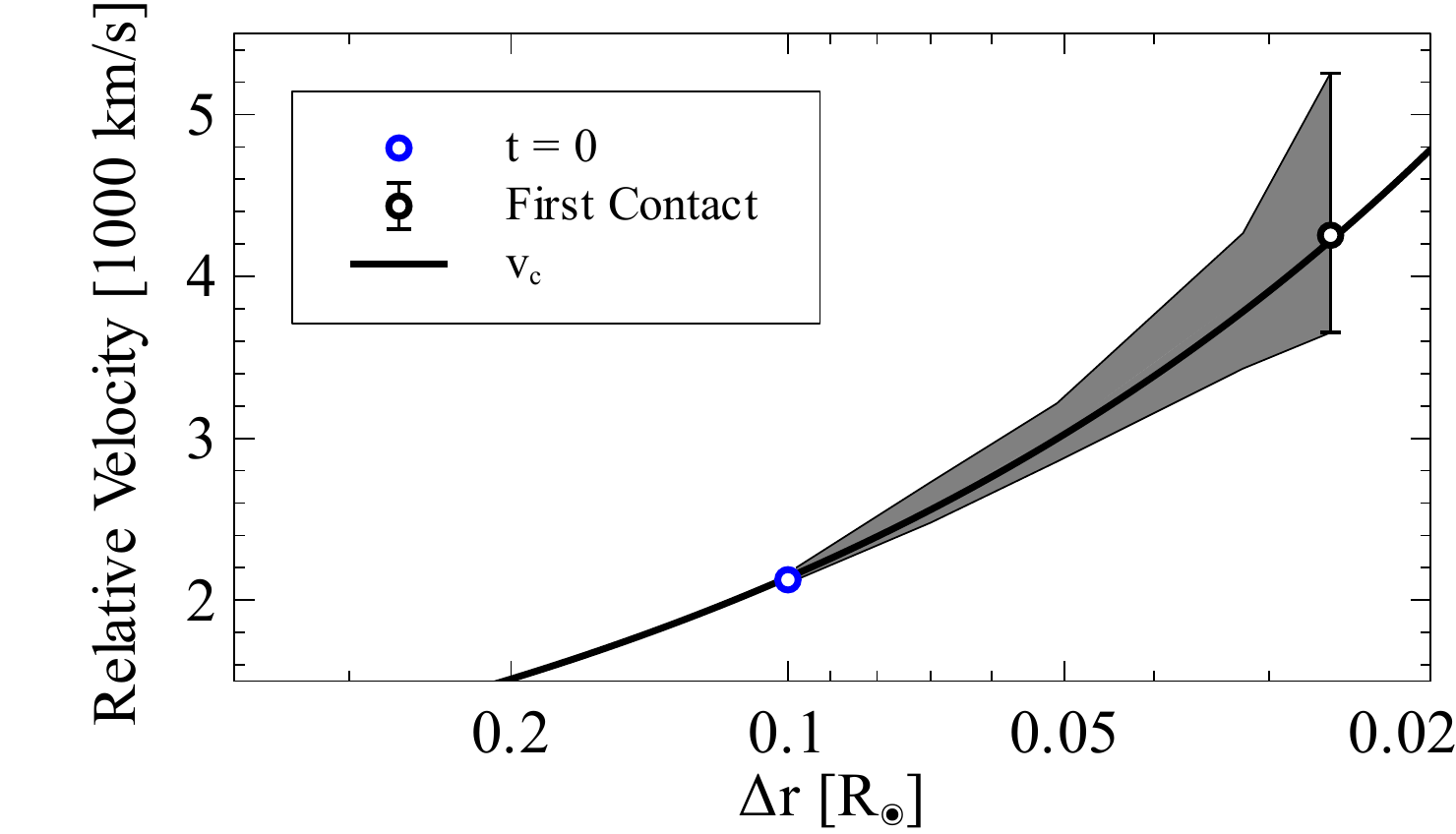}
\caption{The velocity evolution from our initial conditions to the moment of first contact, indicating strong velocity gradients induced by tidal forces. The shaded area denotes the spread in relative $x$-velocities, and $v_{\rm c}$ is the relative velocity of the centers of mass.}
\label{fig:rk}
\end{figure}

Figure \ref{fig:rk} compares the initial conditions of the 0.64\msol$\times2$, head-on collision to the velocity gradient that is introduced by tidal forces. The relative velocity of the centers of mass can be predicted analytically for a zero impact parameter collision with $v_{\rm c}=\left[{2G(M_1+M_2)}/{\Delta r}\right]^{1/2}$, where $M_i$ are the masses of the constituent white dwarfs and $\Delta r$ is the separation of their centers of mass. 

As Figure \ref{fig:rk} demonstrates, when the stars are allowed to free-fall in SNSPH from larger separations, such as the separation of 0.1R\sol used throughout this paper, the velocity gradients that arise from tidal distortions are non-negligible. As will be shown in \S3.1.2, the magnitudes of these velocities and their spreads play an important role in the final outcomes and the \nickel[56] yields of each simulation as they determine how much kinetic energy is converted to thermal energy, and thusly, when carbon-ignition occurs. A shooting-method was used to determine the necessary, initial vertical separation that resulted in the final impact parameter that we desired at the moment of impact. Initial velocities for all of our collision scenarios with zero impact parameter are given in Table \ref{table:ic}. 

\begin{table*}[ht]
\caption{\rm{Initial velocities of each component star in the head-on cases of each mass pair for initial separations of 0.1R\sol. All velocities are relative to the center of mass.}}
\centering
\begin{tabular}{c | c c | c c}
\hline\hline
\# & $m_1$ [\msol] & $m_2$ [\msol] & $-v_1$ [$\times10^3$km/s] & $v_2$ [$\times10^3$km/s]\\
\hline
1 & 0.64 & 0.64 & 1.10 & 1.10\\
2 & 0.64 & 0.81 & 1.31 & 1.03\\
3 & 0.64 & 1.06 & 1.58 & 0.95\\
4 & 0.81 & 0.81 & 1.24 & 1.24 \\
5 & 0.81 & 1.06 & 1.51 & 1.15\\
6 & 0.96 & 0.96 & 1.35 & 1.35\\
7 & 1.06 & 1.06 & 1.41 & 1.41 \\
8 & 0.50 & 0.50 & 0.97 & 0.97\\
\hline
\end{tabular}
\label{table:ic}
\end{table*}

Likewise, all of our stars are initialized with 50\% \carbon[12] and 50\% \oxygen[16] throughout. This approximates typical carbon-oxygen white dwarf compositions, and we use the Helmholtz free-energy EOS (Timmes \& Arnett 1999; Timmes \& Swesty 2000). 

\subsection{Hybrid Burner}

Most large, hydrodynamic codes use some form of a hydrostatic nuclear network (\eg Eggleton 1971; Weaver \etal 1978; Arnett 1994; Fryxell \etal 2000; Starrfield \etal 2000; Herwig 2004; Young \& Arnett 2005; Nonaka \etal 2008). That is, the thermodynamic conditions present at the start of a burn calculation are not altered until the next hydrodynamic time step, which often times is controlled by abundance or energy changes from the burn calculation rather than a pure Courant condition. The effect of this is to fix the temperature-dependent reaction rates throughout the hydrodynamic time step to what they were at its start. 

\begin{figure}[ht]
\centering
\includegraphics[width=6.25cm,height=4.17cm]{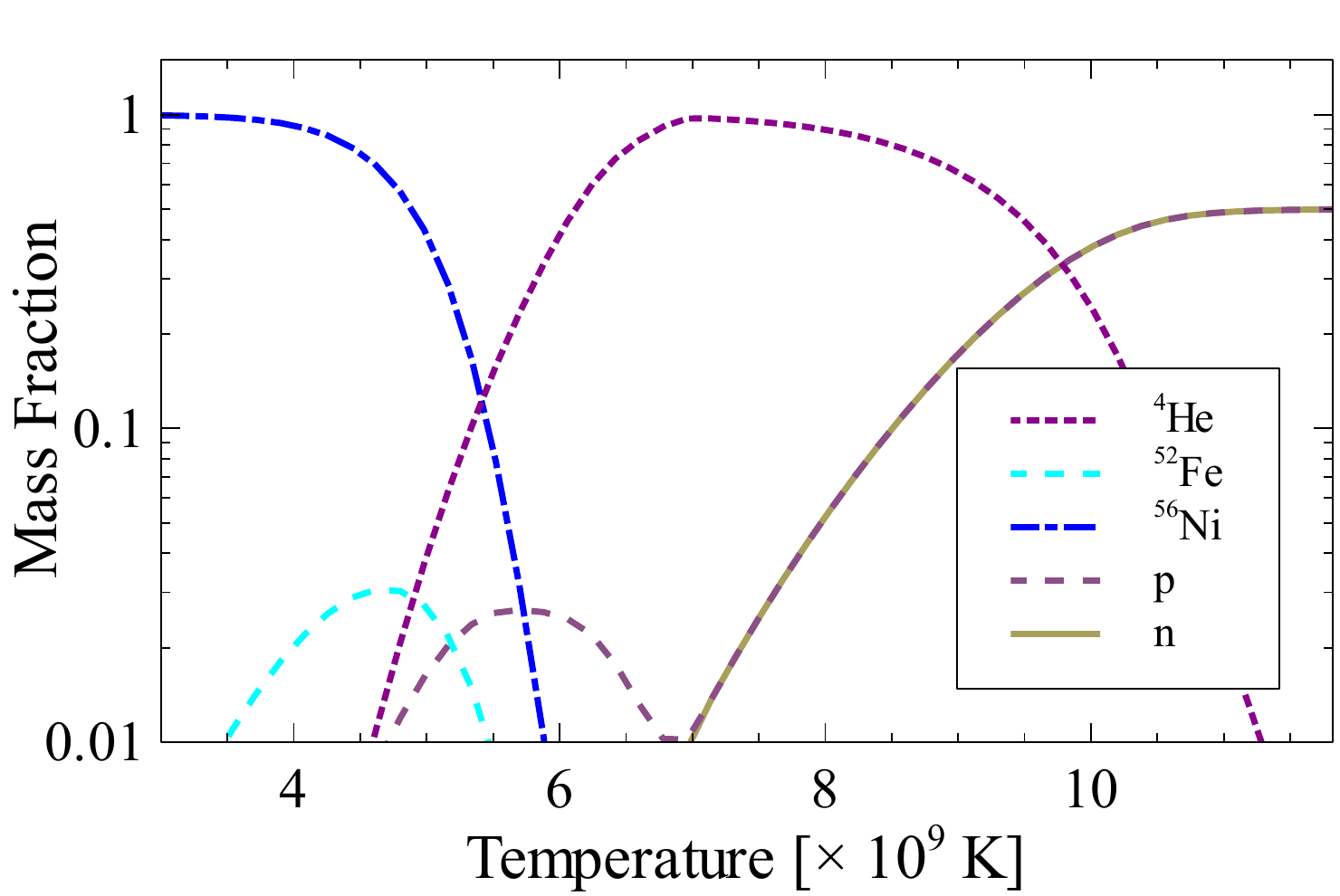}
\caption{NSE distributions for $\rho=$1e7 g cm$^{-3}$ and Y$_e$=0.5 in an $\alpha$-centric nuclear network. Proton and neutron mass fractions are plotted for reference. At $T_9 \approx 6$, $^4$He begins to dominate the isotope distribution.}
\label{fig:nse}
\end{figure}

There are several reasons for running a simulation this way, the most important of which is to avoid a decoupling of the nuclear network from the hydrodynamic calculation. However, for limited spatial, mass and time resolutions, this approximation - that the thermodynamic conditions do not change rapidly enough during a burn to warrant a sub-cycle recalculation of the nuclear reaction rates - fails in regimes where the nuclear reactions are strongly temperature dependent, such as at temperatures where photo-disintegration is the dominant nuclear process. As Figure \ref{fig:nse} shows below, the nuclear statistical equilibrium (NSE) state for material with $\rho = 1\times 10^6$ g cm$^{-3}$ at $T_9 \approx 7$ has most of the heavy isotopes photo-disintegrating back to $^4$He.

In such a regime, the material undergoing photo-disintegration experiences what amounts to an abrupt phase change through a strongly endothermic reaction. In nature, this reaction should rapidly cool the material before complete photo-disintegration, allowing these liberated $\alpha$-particles to react with other isotopes. However, a hydrostatic burn will overestimate the time-scale for this cooling as it assumes a full hydrodynamic time step is necessary for relevant pressure or temperature changes. With the temperature remaining fixed over an artificially long time, this approach results in the nuclear network removing far too much internal energy, $u$, to be physical. 

Typically, one attempts to limit the impact of such a phase change by relying on a global time-step minimization scheme of the form
\be
\Delta t_{n+1}=\min\left[\Delta t_c,\Delta t_{n}\times f_u\times\left(\frac{u^i_{n-1}}{u^i_{n}-u^i_{n-1}}\right)\right],
\label{eq:min}
\ee
where the subscript $n$ refers to the iteration number, $\Delta t_c$ is the Courant time, $u^i$ is the specific internal energy of the $i^{\rm th}$ particle, and $f_u$ is a dimensionless parameter which constrains the maximum allowable change in energy. Our global time-step is also controlled in this manner. In practice, the conditions immediately prior to photo-disintegration will fix the next time-step, $\Delta t_{n+1}$, to of order $10^{-5}$s for $f_u=0.30$. However, even this time-step is too large to capture the relevant temperature changes effecting the reaction rates on time-scales of order $10^{-12}$s.

The alternative approach to a hydrostatic burn is to use a ``self-heating/cooling" nuclear network that simultaneously integrates an energy equation and the abundance equation self-consistently (see \eg M\"uller 1986). When applied to a particle code like SPH, this type of calculation keeps $\rho$ fixed, but updates temperature in a fashion that is consistent with the equation of state and the new internal energy at each sub-cycle. 

The ordinary differential equation that governs a hydrostatic burn calculation of the abundance of an isotope $Y_i$, assuming the mass diffusion gradients are negligible, is of the form
\ba
\dot{Y_i}&=&\sum_jC_iR_jY_j\nonumber\\
&+&\sum_{j,k}\frac{C_i}{C_j!C_k!}\rho N_AR_{j,k}Y_jY_k\nonumber\\
&+&\sum_{j,k,l}\frac{C_i}{C_j!C_k!C_l!}\rho^2N_A^2R_{j,k,l}Y_jY_kY_l,
\label{eq:ydot}
\ea
where the coefficients $C_{i..l}$ specify how many particles of the $i^{\rm th}$ species are created or destroyed, and $R_{i..l}$ are the temperature-dependent reaction rates for each of the different reaction types. The first term describes weak reactions ($\beta$-decays and electron captures) and photo-disintegrations, the second describes two-body reactions of the type \carbon[12]($\alpha$,$\gamma$)\oxygen[16], and the third term describes three-body reactions, such as \helium[4](2$\alpha$,$\gamma$)\carbon[12]. The energy generation ODE takes the form
\be
\dot{\epsilon}=-N_A\sum_i\dot{Y_i}m_ic^2,
\label{eq:edot}
\ee
where $m_i$ is the rest-mass of the $i^{\rm th}$ isotope (see \eg Benz \etal 1989b). The density and temperature equations are simply $\dot{\rho}=0$ and $\dot{T}=0$, respectively, in a hydrostatic burn. 

A self-heating at constant density calculation modifies only the temperature equation, starting from the first law of thermodynamics in specific mass units,
\be
\frac{du}{dt}-\frac{P}{\rho^2}\frac{d\rho}{dt}=T\frac{ds}{dt},
\ee
where $du/dt$ is the change in specific energy and $ds/dt$ is the change in specific entropy. In accordance with $\dot{\rho}=0$ and employing the identity $T\dot{s}=\dot{\epsilon}$, this reduces to
\ba
\frac{\partial u}{\partial T}\frac{dT}{dt}&=&\dot{\epsilon}\nonumber\\
\dot{T}&=&\frac{\dot{\epsilon}}{c_{\rm v}},
\label{eq:selfheat}
\ea
where $c_{\rm v}$ is the specific heat capacity at a constant volume. Equations (\ref{eq:ydot}), (\ref{eq:edot}), and (\ref{eq:selfheat}) are evolved simultaneously and self-consistently (M\"uller 1986). 

\begin{figure}[ht]
\centering
\includegraphics[width=8cm,height=6cm]{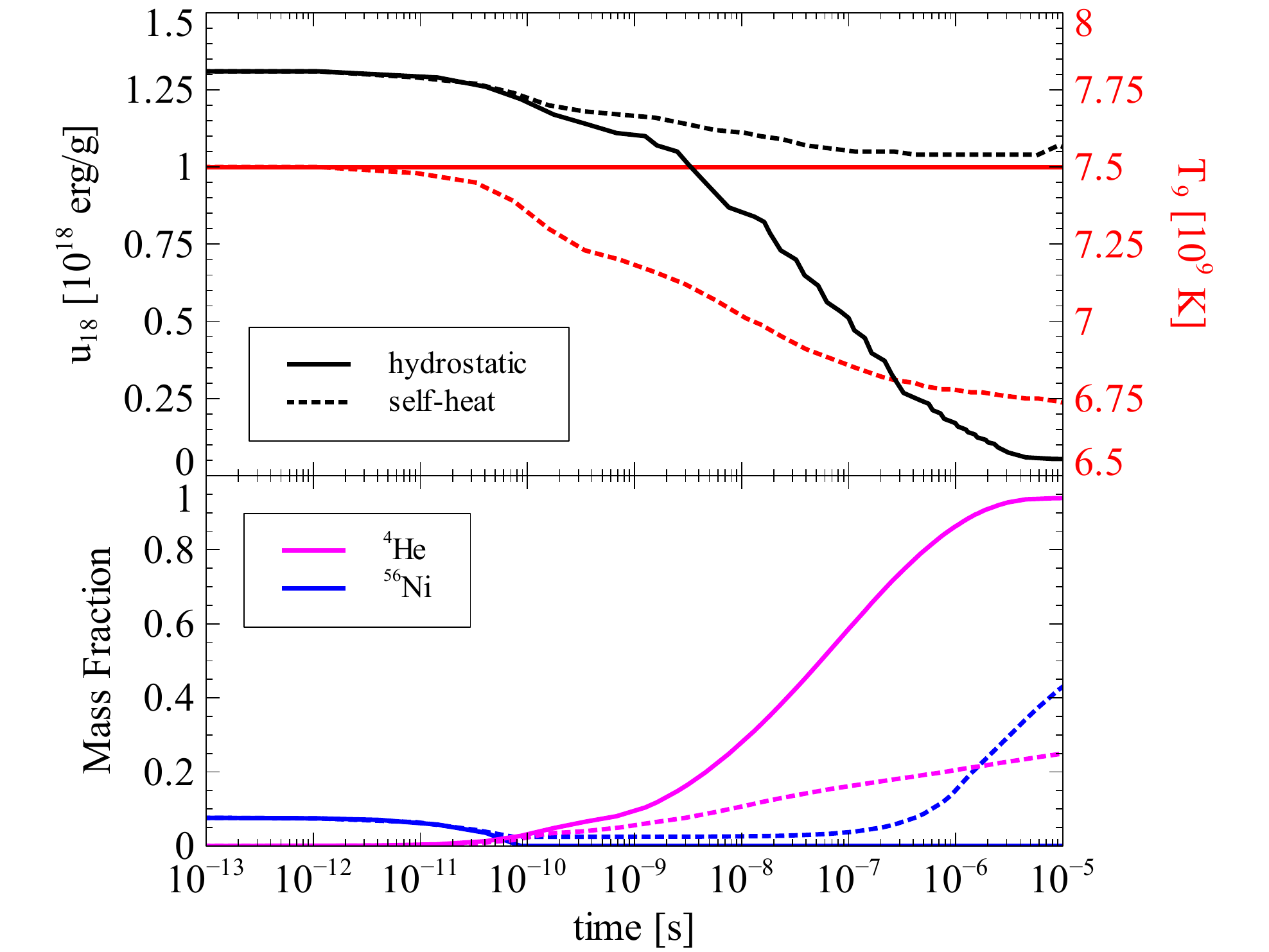}
\caption{Calculations of the energy, temperature, and composition of a particle after a representative time-step as determined by Equation (\ref{eq:min}) with $f_u=0.30$. Solid lines show the implicit integrations from a hydrostatic calculation, while dotted lines indicate those for a self-heating calculation. In both cases, $\rho$ is kept constant, while in the self-heat calculation, the temperature, and thus the nuclear reaction rates, are recalculated at each implicit integration step, consistent with the first law of thermodynamics.}
\label{fig:hybrid}
\end{figure}

At very high spatial resolutions and small time-steps, the self-heating approach would be identical to the hydrostatic approach. As Figure \ref{fig:hybrid} shows for the energy, temperature and composition of a single particle over a finite and relatively large time-step as determined by Equation (\ref{eq:min}) with $f_u=0.30$, these two burning calculations reach very different conclusions about the final energy and composition of the particle after photo-disintegration. 

To capture the relevant temperature changes using a hydrostatic approach would require a global time-step $\sim10^{-12}$s, where the temperatures of the two calculations have diverged by $\approx5\%$. This is problematic for two reasons: 1) such a time-step cannot be predicted from the conditions immediately prior to photo-disintegration, and 2) having such a small global time-step exceeds the limit of machine precision for many hydrodynamic codes, SNSPH included. When SNSPH attempts to calculate velocities for the next time-step using $\Delta t\sim10^{-12}$s, it often fails or returns zero.

Unfortunately, a self-heating nuclear network can also expose the weakness of a mass resolution limit. For a typical simulation of $10^6$ particles, each particle has a mass of $\approx 10^{27}$g. A self-heating nuclear calculation for carbon-burning of such large masses becomes rapidly explosive on time-scales approaching the Courant limit. Without any mechanism for energy transport on such short time-scales, the assumption of a homogeneous burn of all $10^{27}$g begins to break down. The vigorous burning of so much material rapidly liberates more energy than the binding energy of the star. 

Our ``middle-path" solution to these two extremes is a hybrid-burning scheme wherein a combination of these two approaches is used under different circumstances. Since the hydrostatic approach is a better approximation for exothermic reactions at our resolution limit, the self-heating/cooling approach is only employed for particles that undergo strong, net endothermic reactions such as photo-disintegration. This allows these particles to smoothly ``step over" the photo-disintegration phase change without artificially losing too much energy. We apply this approach, along with the time-step minimization of Equation (\ref{eq:min}), to the $\alpha$-chain {\it aprox13} nuclear network (Timmes 1999; Timmes \etal 2000) by imposing the condition that if $\dot{\epsilon}<0$ after a hydrostatic burn, the burn is recalculated employing Equation (\ref{eq:selfheat}).

While stepping through a photo-disintegration process is an interesting wrinkle for numerical simulations of double-degenerate white dwarf collisions, it is not a significant factor for \nickel[56] production. In all of our simulated cases, we found that on average $<2$\% of particles experienced this phase change. For the most part, the local conditions for a number of particles wherein a single particle might undergo photo-disintegration are sufficiently high-energy that the neighboring particles will have already initiated a detonation. Based on the results of our simulations, we do not expect that collisions with yet higher kinetic energies than those attempted here would paradoxically yield less \nickel[56] due to photo-disintegration effects.

\section{Results \& Analysis I - Detonations}

Our previous work narrowed the range of pertinent impact parameters to three scenarios, which we revisited for each of our mass combinations. We simulated head-on impacts, partially grazing collisional impacts, and fully grazing/glancing impacts. Table \ref{table:runs} summarizes the \nickel[56] yields of each of our simulations. In this table, the impact parameter, $b$ (the vertical separation between the cores of both white dwarfs at the moment of impact) is given as the fraction of the radius of the primary white dwarf. Thus the $b=0$ column shows the yields for head-on impacts, the $b=1$ column indicates a full white dwarf radius and $b=2$ indicates 2 white dwarf radii, or a fully grazing impact.

\begin{table*}[ht]
\caption{\rm{Simulation \nickel[56] yields for various mass combinations and parameters. Values in bold are super-Chandrasekhar masses, and values indicated with a (\currency) are those simulations that resulted in remnants. Dashes (--) indicate combinations of parameters we did not simulate. All simulations listed here used $f_u=0.30$ and 200k particles. }}
\centering
\begin{tabular}{c | c c | c | c c c}
\hline\hline
\# & $m_1$ [\msol] & $m_2$ [\msol] & $m_{tot} [\msol] $ & $b=0$ & $b=1$ & $b=2$\\
\hline
1 & 0.64 & 0.64 & 1.28 & 0.51 & 0.47 & \currency\\  
2 & 0.64 & 0.81 & \bf{1.45} & 0.14 & 0.53 & \currency\\
3 & 0.64 & 1.06 & \bf{1.70} & 0.26 & \currency & \currency\\
4 & 0.81 & 0.81 & \bf{1.62} & 0.84 & 0.84 & 0.65\\
5 & 0.81 & 1.06 & \bf{1.87} & 0.90 & 1.13 & \currency\\
6 & 0.96 & 0.96 & \bf{1.92} & 1.27 & 1.32 & 1.33\\
7 & 1.06 & 1.06 & \bf{2.12} & \bf{1.71} & \bf{1.72} & \bf{1.61}\\
8 & 0.50 & 0.50 & 1.00 & 0.00 & -- & --\\
\hline
\end{tabular}
\label{table:runs}
\end{table*}


Dursi \& Timmes (2006) examined the shock-ignited detonation criteria for carbon in a white dwarf using numerical models. They derived a relationship between the density of the carbon fuel and the minimum radius of a burning region that will launch a detonation. For a carbon abundance of $X_{^{12}{\rm C}}=0.5$ and densities typically found in the white dwarfs used in our simulations, $\rho \sim 10^7$g cm$^{-3}$, their results suggest a minimum burning region, or match head size of $r_b\sim10^4$cm. This is three orders of magnitude smaller than our smallest particle size, and properly resolving this criterion would require $\sim10^{12}$ particles. Such a high-resolution study is too expensive with our current computing resources, and therefore, we acknowledge that we cannot resolve the precise detonation mechanism in our simulations. 

The criteria established in Dursi \& Timmes (2006) would suggest that a single particle in any of our simulations can initiate a detonation. However, in order for a detonation to be sustained, the energy that the initiating particle deposits in its neighbors must be sufficient to cause those neighbors to liberate an equal amount of nuclear energy. This somewhat softens the ability of a single particle to initiate a {\it sustainted} detonation. Indeed, in all of our simulations, we found that several particles ignited nearly simultaneously, or at least outside of causal contact with one another in order to initiate a sustained detonation. Moreover, the pressure gradient established by particles neighboring those that reached ignition needed to be favorable for a significant and rapid energy deposition.

In SPH, energy is shared between particles via $PdV$ work with
\be
\dot{u}_{ij}= \frac{P_i}{\rho_i^2}m_j \Delta v_{ij} \cdot \nabla_i W_{ij},
\ee
where $P_i$ and $\rho_i$ are the pressure and density of particle $i$, $m_j$ is the mass of particle $j$, $\Delta v_{ij}$ is the difference in velocities of particles $i$ and $j$, and $W_{ij}$ is the SPH smoothing kernel. For each particle $i$, there is an implied sum over all particles $j$. In SPH formalism, the condition for a sustained detonation would require that this quantity is large enough to ignite explosive burning in particle $i$. Put another way, if particle $i$ generates energy $\epsilon_i$ at time $t_0$, $\dot{u}_{ji}$ must be sufficient such that at time $t_0+\Delta t$, $\dot{\epsilon}_j\approx\dot{\epsilon}_i$, where $\Delta t$ is the Courant time. This requires a proportionality between the energy generation rate in particle $i$ and the pressure gradient with its nearest neighbors, and to first order, this criterion reduces to
\be
\dot{\epsilon_i}  \ge \frac{P_i}{c_s\rho_i^2}\nabla P_{ij}.
\ee

For a given energy generation rate, large and positive pressure gradients can inhibit a detonation breakout. In situations where particles ignited carbon-burning, but were nevertheless unable to deposit enough energy into their neighbors to cause them to also ignite, the material settled into a slow-burn regime rather than detonating. While we cannot resolve the detonation mechanism to the desired precision, we compared one-dimensional ZND detonation profiles (see \eg Fickett \& Davis 1979) with detonation profiles from one-dimensional slices through the SNSPH models and concluded that our collision calculations are resolving the detonation widths and structures to within 20\%.

\subsection{Mass Pair 1 - 0.64$\msol\times2$}
\subsubsection{Fiducial Case}

We recalculated the $\approx 0.6$\msol equal mass case as in Raskin \etal (2009) to establish a baseline comparison with our equal-mass particle configuration and hybrid-burner technique. Empirical white dwarf mass functions (\eg, Williams, Bolte, \& Koester 2004) suggest collisions with this mass pair are expected to be among the most common. With our equal-mass particle constraint, the final mass of the star used in the simulations came to 0.64\msol.

\begin{figure*}
\centering
\includegraphics[width=11.25cm,height=9.0cm]{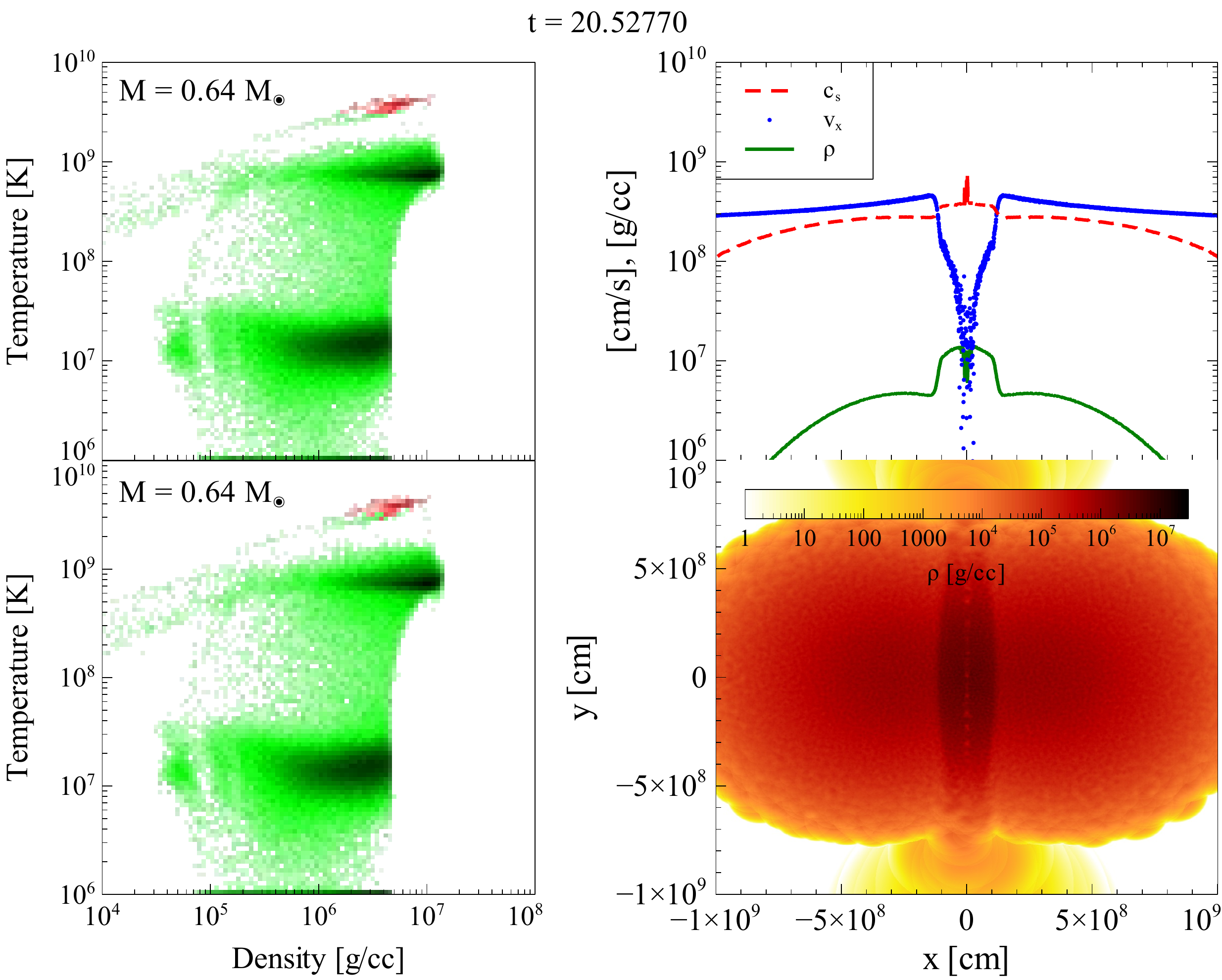}
\caption{\textbf{Left Panels:} Density vs. temperature for all particles from each constituent star in mass pair 1, 0.64\msol$\times2$ and $b=0$. Each point is colored indicating the isotope(s) with the greatest abundance and by the particle number density at each $\rho-T$ coordinate. Green indicates high concentrations of carbon and oxygen, red indicates silicon group elements, and blue indicates iron-peak elements, most predominantly \nickel[56]. The darker the color in each group, the higher the particle number density. \textbf{Right-top Panel:} Sound speed, infall velocity, and density for particles lying on the $x$-axis. \textbf{Right-bottom Panel:} A slice in the $x$-$y$ plane of particle densities.}
\label{fig:1450}
\end{figure*}

The right-top panel of Figure \ref{fig:1450} shows that when the stars first collide the infall speeds, $v_x$ of material entering the shocked region are greater than the sounds speeds, $c_s$, resulting in a stalled shock in that region. The conditions in the center plane of this shocked region (the $y$-$z$ plane, $\rho \approx 10^{6.5-7}$ g cm$^{-3}$ and $T_9\approx 1$) are sufficient to ignite carbon with an energy-generation rate scaling roughly as $\dot{\epsilon} \sim \rho T^{22}$, burning up to silicon at $T_9\approx 3$. The separation of material into three distinct phases is clearest in the left two panels of Figure \ref{fig:1450}, which plots particle number density in the $\rho$-$T$ plane. The lower, more populated region is unshocked, carbon-oxygen material and is indicated in green. The less populated middle region, also represented with green at $T_9\approx 1$, is shocked material that has yet to reach the critical conditions for carbon ignition, and the upper, sparse region is material that has begun burning carbon to silicon-group elements, represented in red. 

The pressure gradient slopes positively in all directions away from the geometric center where this early burning begins, which is in fact at lower densities than the surrounding shocked medium. While the whole of the shocked region continues to heat up, causing more material to ignite near the center, the steep pressure gradient prevents the energy liberated by burning to initiate a detonation. For these burning particles, with silicon ash behind them and higher pressure carbon in front, the energy they deposit in their neighbors is insufficient to greatly alter their energy-generation rate. Instead, the burning region grows only as fast as material is heated to the critical temperatures needed for carbon ignition, $T_9 \approx 1$, by the conversion of kinetic energy to thermal energy.

\begin{figure*}
\centering
\includegraphics[width=11.25cm,height=9.0cm]{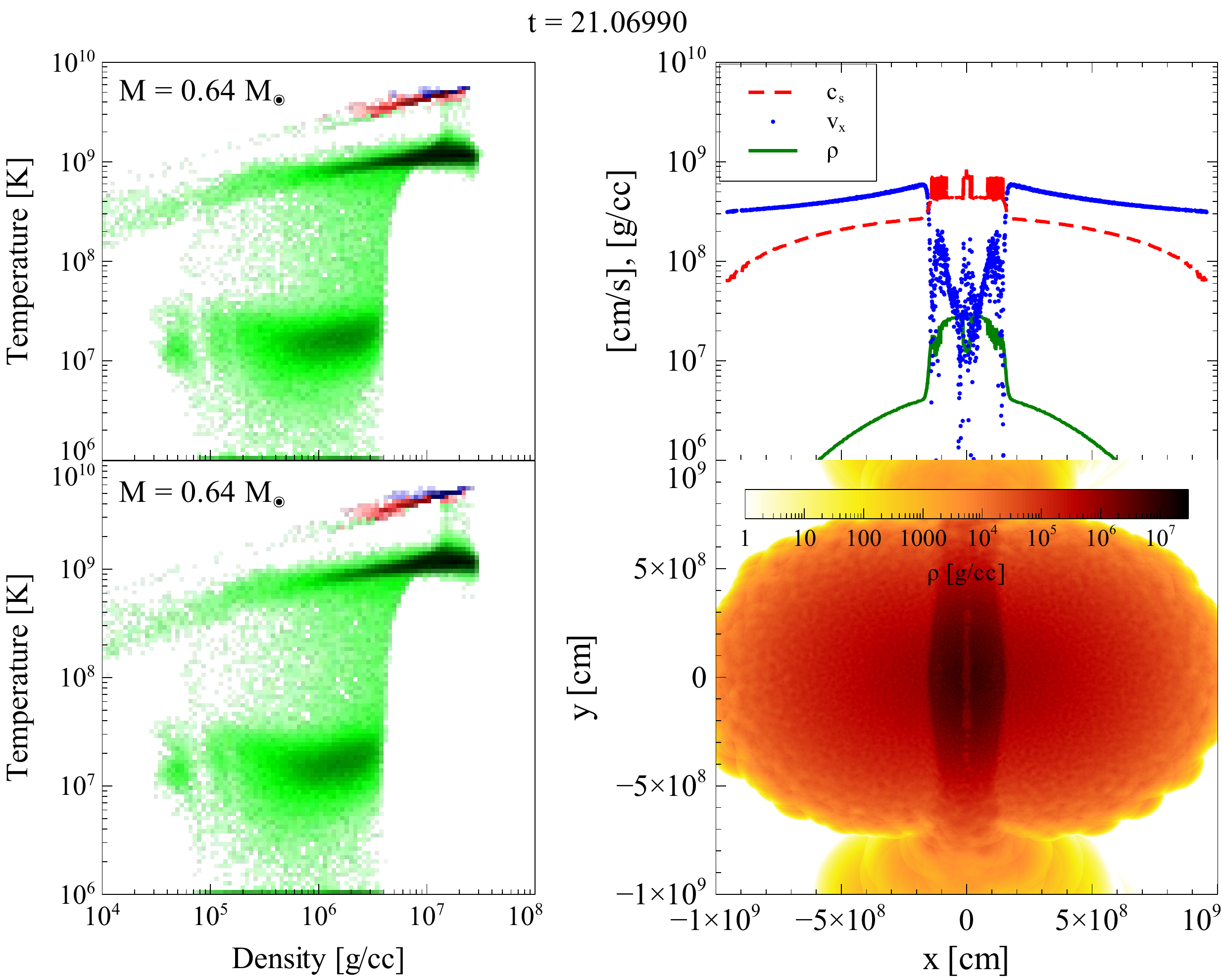}
\caption{Same format as Figure \ref{fig:1450}, at a later time in the simulation.}
\label{fig:1875}
\end{figure*}

Approximately two seconds after the stars first collide, sufficiently high temperatures and densities are reached at the edges of the shocked region to initiate carbon-burning. In these locations, the pressure gradient slopes negatively in all directions. The liberated energy is free to break out, initiating detonations at the ignition points. The sound speeds in these zones are raised higher than the infall speeds due to the rapid rise in temperature, and \nickel[56] begins to appear in large quantities, indicated in blue in the left panels of Figure \ref{fig:1875}.

\begin{figure*}
\centering
\includegraphics[width=11.25cm,height=9.0cm]{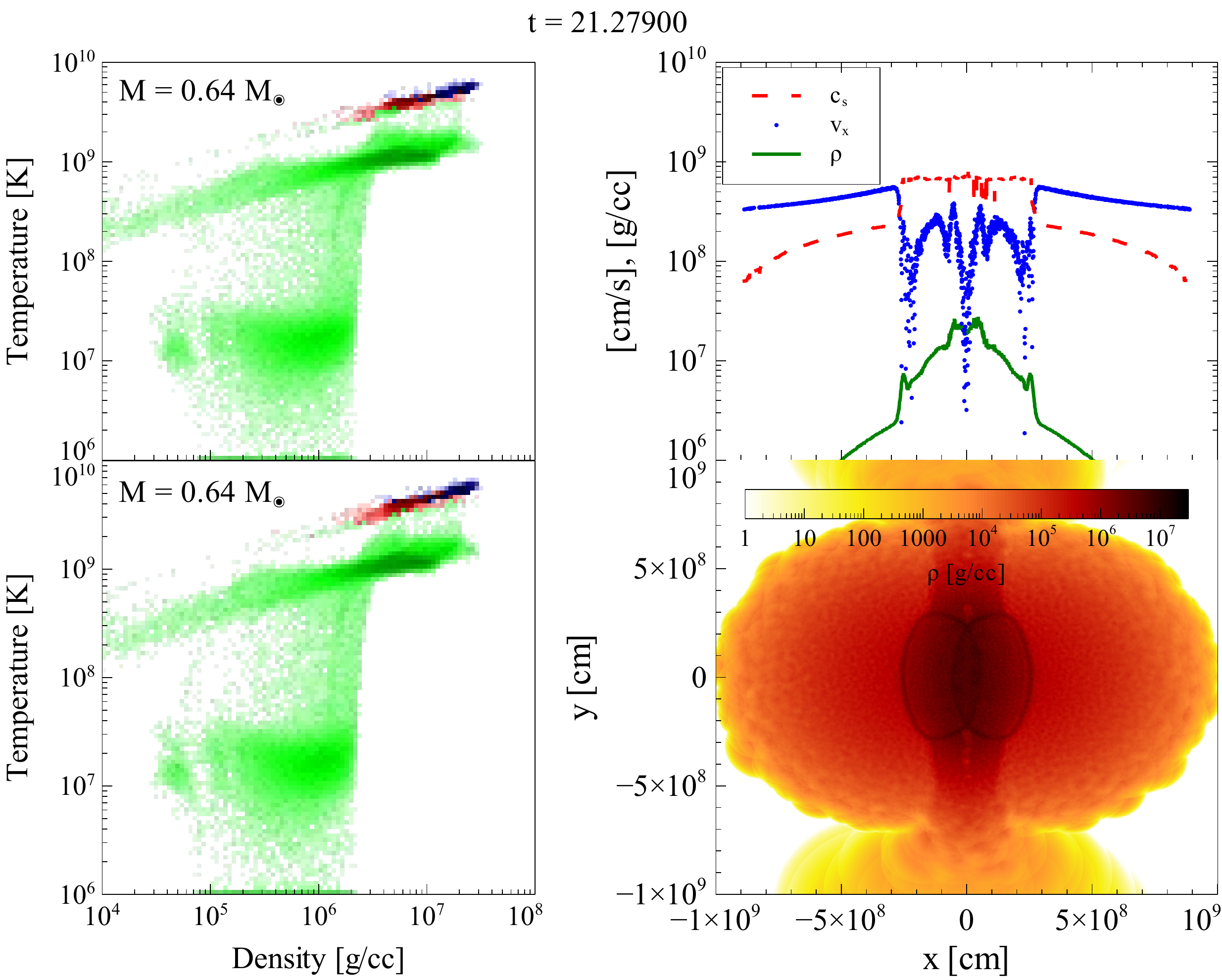}
\caption{Same format as Figure \ref{fig:1450}, at a later time in the simulation.}
\label{fig:3725}
\end{figure*}

\begin{figure*}
\centering
\includegraphics[width=11.25cm,height=9.0cm]{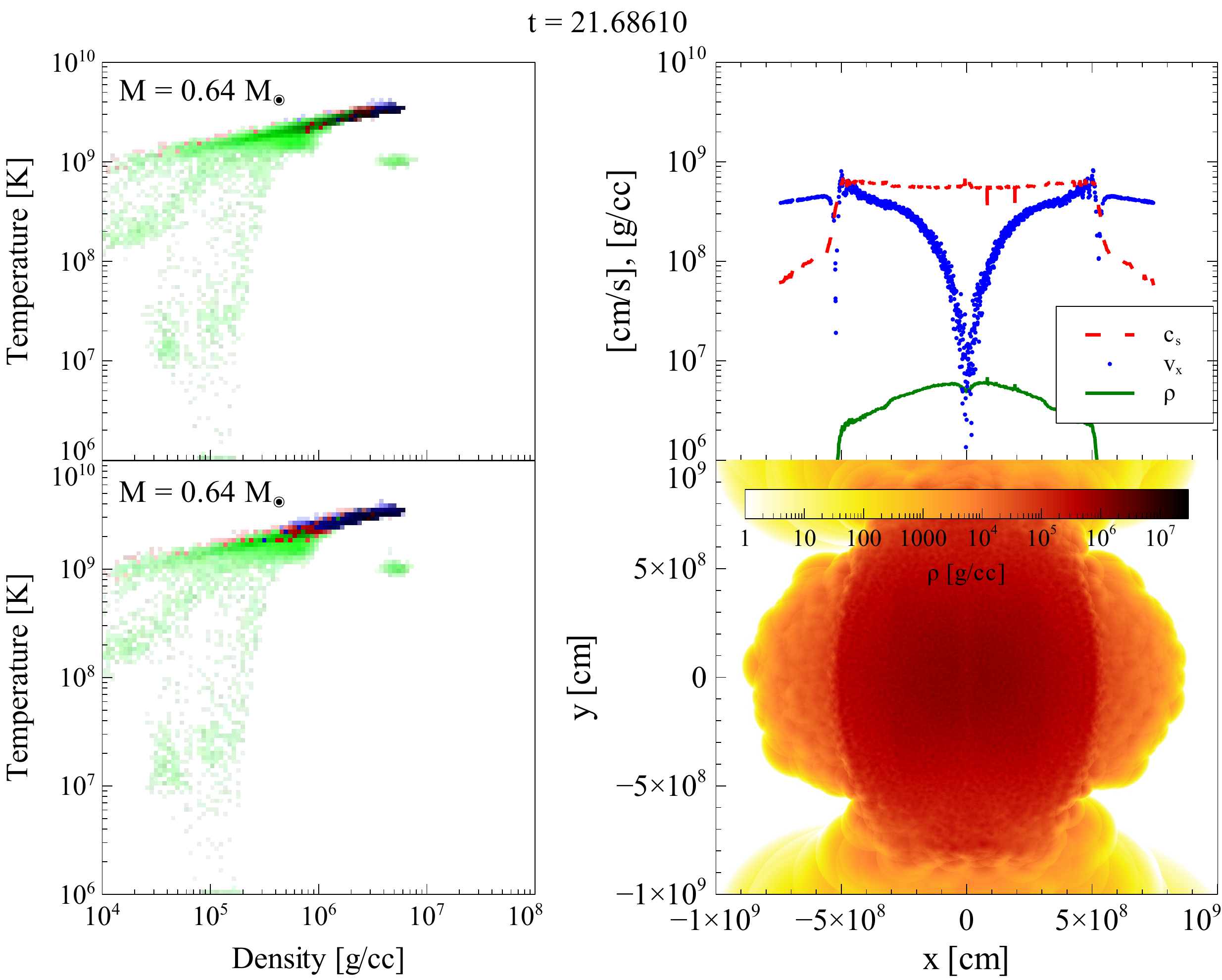}
\caption{Same format as Figure \ref{fig:1450}, at a later time in the simulation.}
\label{fig:4275}
\end{figure*}

Sustained detonation fronts then propagate through the unburned material, as well as the silicon ``ash'' that lies in the shocked region. As Figure \ref{fig:3725} shows, significantly more \nickel[56] is produced during this phase. In Figure \ref{fig:3725}, it is also evident that the shocks overtake one another inside the contact zone, shocking the material a second time and producing yet more \nickel[56]. Less than one second after the detonations began, the entire system has become unbound, freezing out the nuclear reactions, as can be seen in Figure \ref{fig:4275}. The final \nickel[56] yield for this simulation was 0.51\msol. 

\begin{figure}[ht]
\centering
\includegraphics[width=7cm,height=7cm]{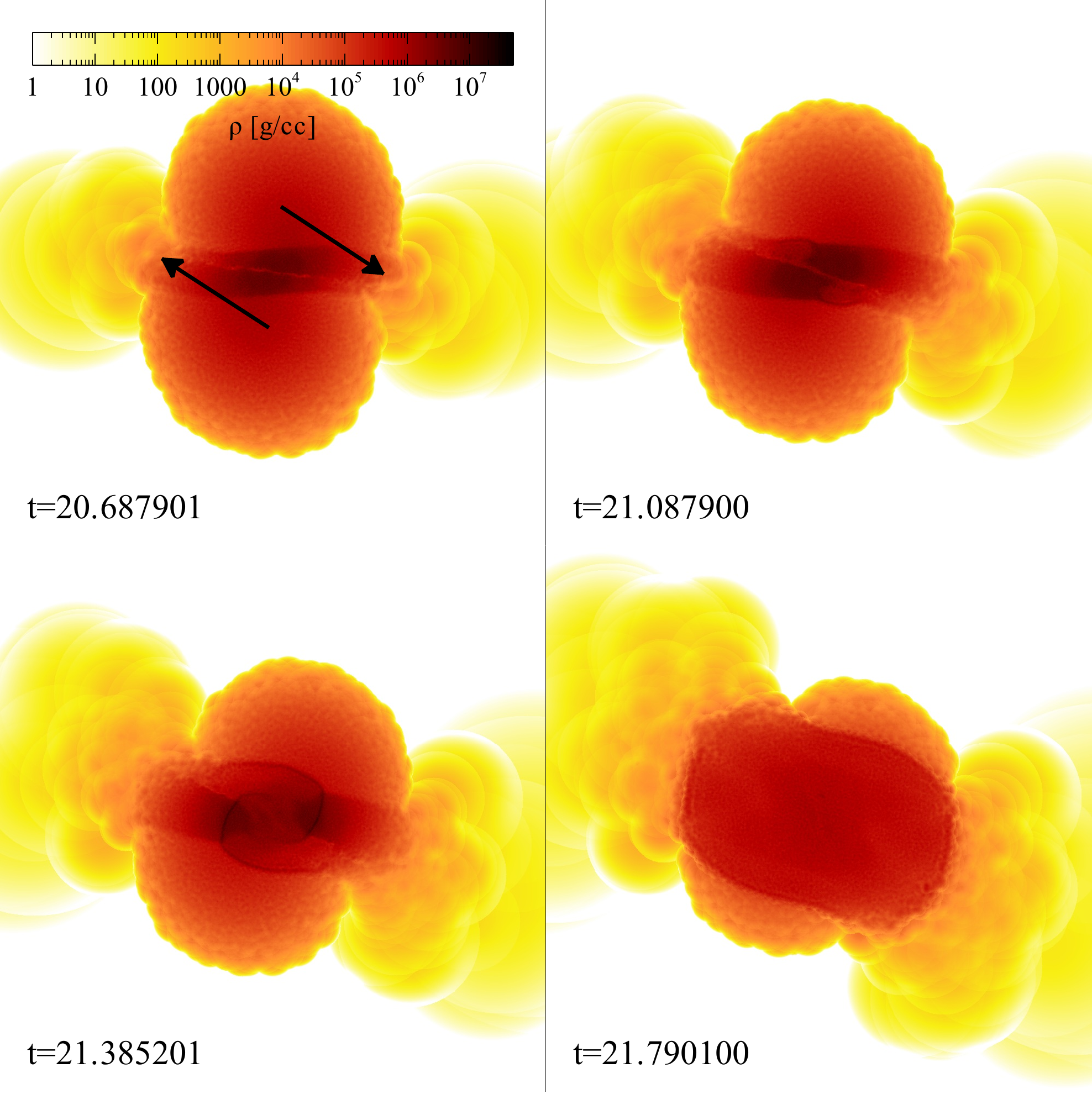}
\caption{A 2D slice of interpolated densities through the $x$-$y$ plane of the $b=1$ case of two 0.64\msol white dwarfs colliding. Four snapshots at different times are shown. Arrows in the top-left panel indicate the directions of motion of each star.}
\label{fig:0p6x2_half}
\end{figure}

In the $b=1$ simulation, the added angular momentum distorted the shocked region between the two stars, resulting in detonations lighting off-center and off-axis as compared to the $b=0$ case. As Figure \ref{fig:0p6x2_half} shows, the detonation waves traveling through the densest portions of the shocked regions where the sound speed is highest, twist the material into a uniquely anisotropic configuration. Moreover, because much of the material is traveling nearly perpendicular to the shock, the density in the pre-detonation, shocked region is lower than in the $b=0$ case. This reduces \nickel[56] production by about 7\% to 0.47\msol.

\begin{figure}[ht]
\centering
\includegraphics[width=5.25cm,height=6cm]{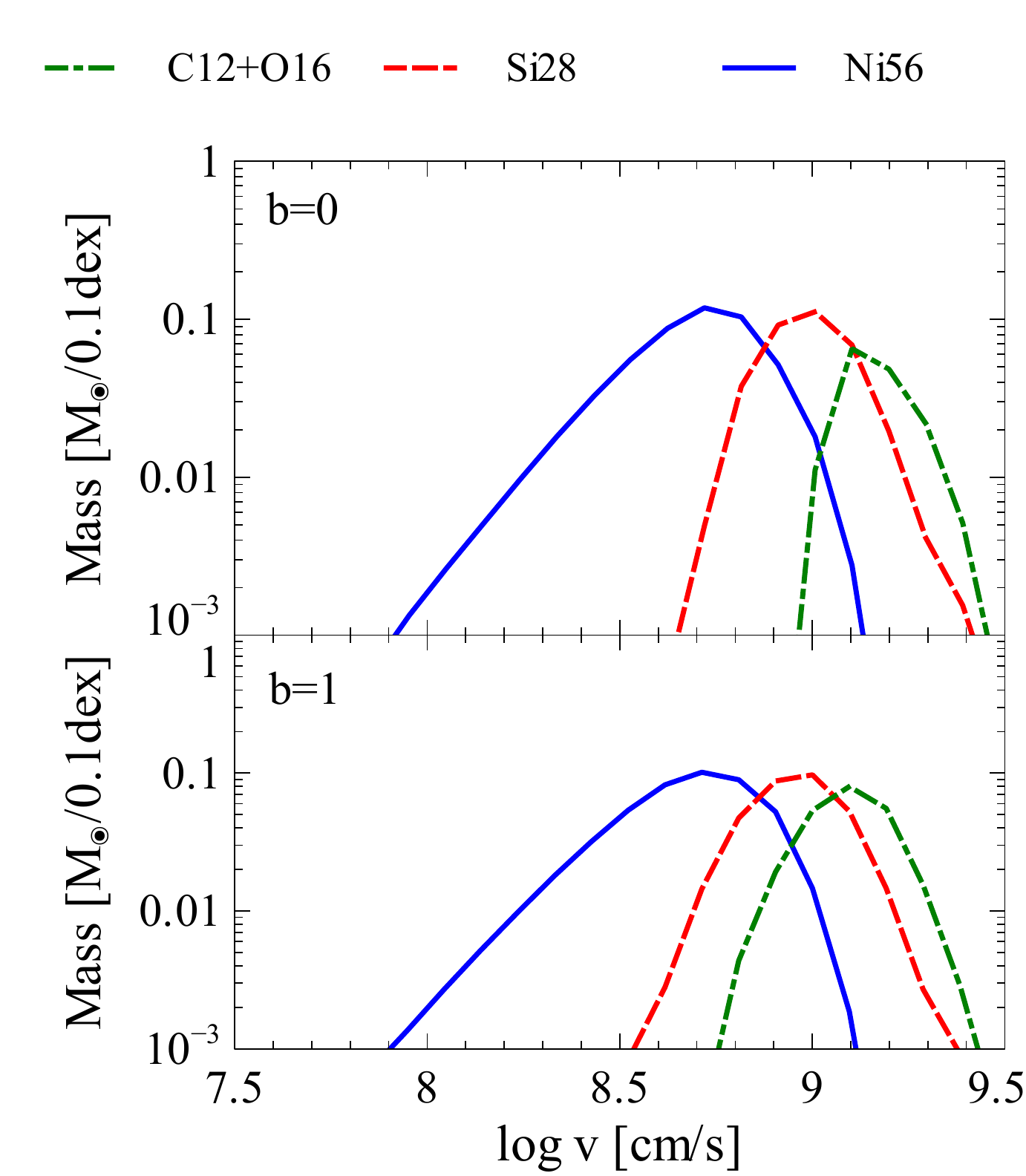}
\caption{Masses of several isotopes at logarithmically spaced velocity bins for the $b=0$ and $b=1$ cases of mass pairing 1, 0.64\msol$\times2$.}
\label{fig:vel1}
\end{figure}

Since the post-explosion, expansion phase is homologous, the pattern of isotopes present at the moment the system becomes unbound is not altered by the expansion. Therefore, the velocities plotted in Figure \ref{fig:vel1} for several isotopes in the $b=0$ and $b=1$ cases are directly related to the radial distribution of the isotopes. 

The velocity structure preserves the isotopic segregation expected behind the burning front, with a progression from complete burning of carbon \& oxygen to iron-peak elements, though silicon-group elements, and finally ending with an unburned or only partially burned carbon \& oxygen envelope. This layered structure is in agreement with the observations of Scalzo \etal (2010) and others of type Ia SNe suspected of having been produced from double-degenerate progenitor scenarios. 

The $b=2$ scenario for this mass pair did not feature a detonation, and instead, resulted in a hot remnant embedded in a disk. Details of this simulation and its outcome will be discussed in \S4.

\subsubsection{Variations on Parameters}

In order to assess the impact of the time-step on the nuclear yields, we compared three simulations of the 0.64\msol$\times2$, $b=0$ case varying the value of $f_u$ in Equation (\ref{eq:min}); one with a value of $f_u=0.50$, another with $f_u=0.30$, and finally, one with $f_u=0.25$. As the results in table \ref{table:tests} show for the \nickel[56] yields, changes in the value of $f_u$ below 0.5 have little discernible impact on the final outcomes.

\begin{table}[ht]
\caption{\rm{\nickel[56] yields for 0.64\msol$\times2$, $b=0$ simulations with variations on the parameter $f_u$ and particle count.}}
\centering
\begin{tabular}{c c c}
\hline\hline
$f_u$ & Particle Count & \nickel[56]\\
\hline
0.50 & $2\times10^5$ & 0.51\\
0.30 & $2\times10^5$ & 0.51\\
0.25 & $2\times10^5$ & 0.51\\
0.30 & $1\times10^4$ & 0.21\\
0.30 & $4\times10^4$ & 0.31\\
0.30 & $4\times10^5$ & 0.49\\
0.30 & $2\times10^6$ & 0.53\\
\hline
\end{tabular}
\label{table:tests}
\end{table}

The detonations on either side of the shocked region are unique to the 0.64\msol$\times2$ mass pairing and the 0.50\msol$\times2$ mass pairing described in \S3.7. This is due, in large part, to the kinetic energy at impact, which is related directly to the infall speed. With greater infall speeds, the shocked region heats sufficiently to initiate a detonation earlier, and the detonations begin nearer to the central region (the $y$-$z$ plane).

We tested this mechanism with a 0.64\msol$\times2$ collision scenario by giving the constituent stars an artificially high infall velocity to reproduce the kinematic energies associated with collisions of larger masses. In that test, the critical temperatures for carbon ignition were reached in locations nearer to the $y$-$z$ plane, but still displaced enough that the pressure gradient was favorable to a detonation. In this case, the \nickel[56] production was actually depressed, resulting in only 0.39\msol, due to an early detonation coupled with altered shock conditions.

We also tested the effect of velocity gradients (tidal distortions) on the final \nickel[56] yield by running a 0.64\msol$\times2$, $b=0$ collision with an initial separation of only 0.048R\sol with the commensurate relative velocities. In that simulation, the \nickel[56] yield was also depressed, slightly, to 0.48\msol. The combination of infall velocity and tidal distortions are evidently critical for \nickel[56] production.

\begin{figure}[ht]
\centering
\includegraphics[width=7cm,height=3.5cm]{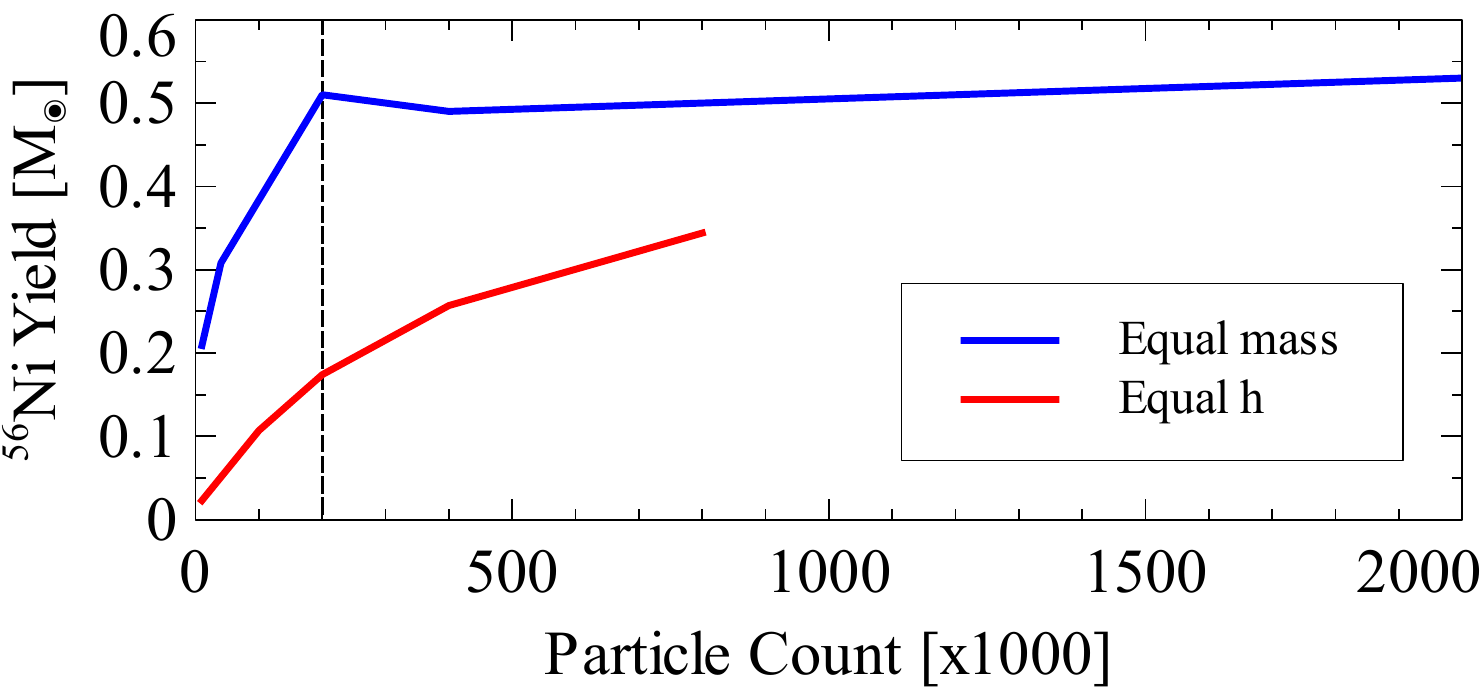}
\caption{Convergence of the \nickel[56] yield with particle count for simulations employing equal mass particles (blue, $0.64\msol\times2$) and equal h particles (red, $0.6\msol\times2$ from Raskin \etal 2009). The dashed, vertical line indicates the number of particles used in simulations throughout this paper.}
\label{fig:convergence}
\end{figure}

However, by far the most important parameter effecting the convergence of the \nickel[56] yield is resolution. We carried out a convergence test of the \nickel[56] yield in mass pair 1, $b=0$, using equal-mass particle setups. We varied particle counts from $10^4$ particles total, to $2\times10^6$. As Figure \ref{fig:convergence} demonstrates, convergence was reached at $2\times10^5$ particles. This compares favorably to previous convergence estimates in Raskin \etal (2009) that concluded $\sim10^6$ particles were needed for convergence using equal-$h$ particle setups. 

Early work on numerical simulations of white dwarf collisions carried out by Benz \etal (1989b) did not have the benefit of modern computational resources to reach these kinds of resolutions. Consequently, the \nickel[56] yields in those simulations were comparatively quite low.

\subsection{Mass Pair 2 - 0.64\msol + 0.81\msol}

\begin{figure*}
\centering
\includegraphics[width=11.25cm,height=9.0cm]{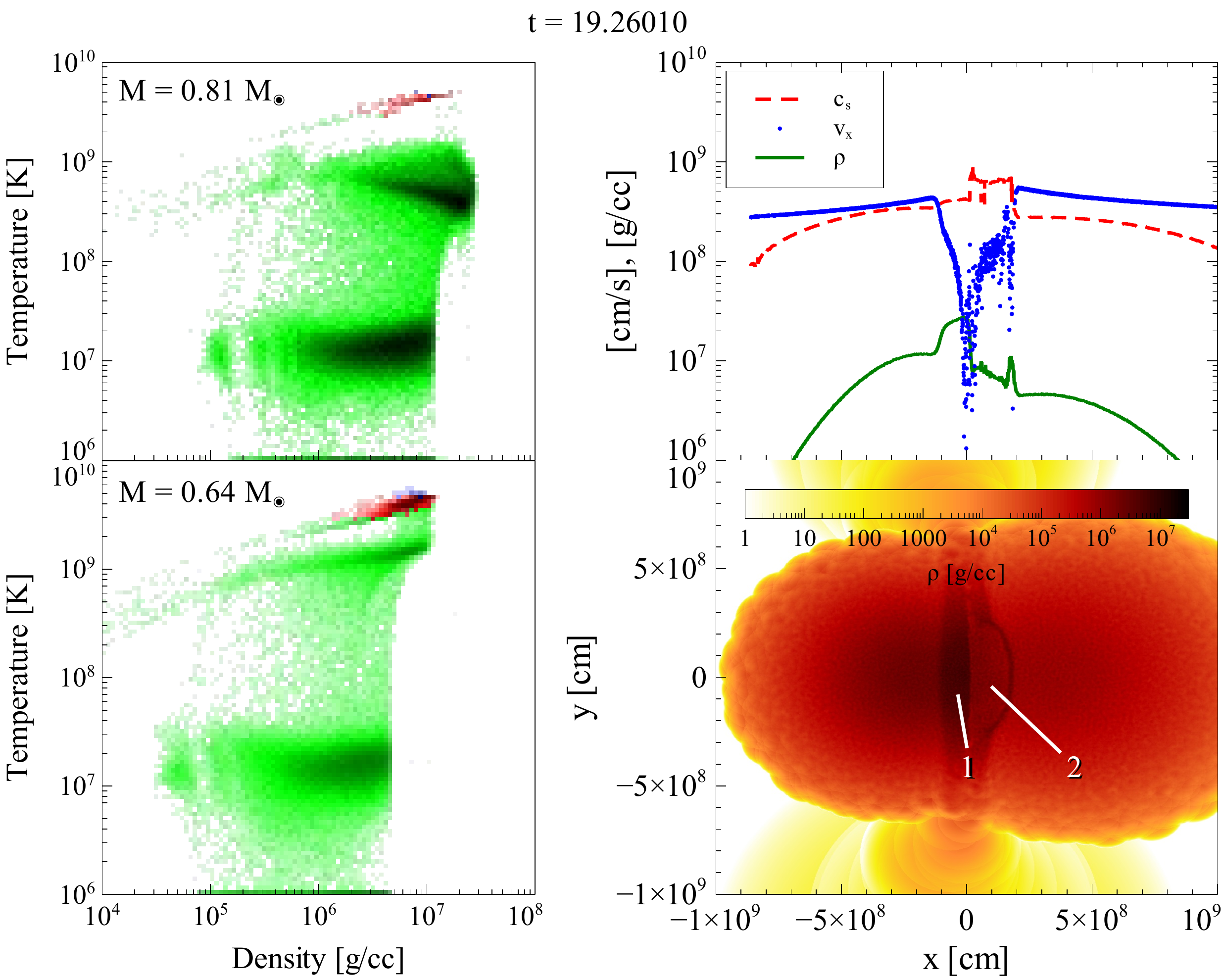}
\caption{Same format as Figure \ref{fig:1450} for mass pair 2, 0.64\msol + 0.81\msol, and $b=0$. Material shocked by the collision is labeled as region 1. Material behind the first detonation shock is labeled as region 2.}
\label{fig:2525}
\end{figure*}

For asymmetrical collisions involving 0.64\msol and 0.81\msol white dwarfs, the higher kinetic energy with which they collide results in several, almost immediate detonations near the center in the $b=0$ scenario. As Figure \ref{fig:2525} shows, these detonation shocks superimpose to form a single, nearly spherical shock front that raises the sound speed above the infall speed for material in the 0.64\msol star, but the pressure gradient leftward of the detonation (region 1) stalls the detonation shock, which can be seen as a higher-density, laminar shock at the rightmost edge of region 1 in Figure \ref{fig:2525}.

\begin{figure*}
\centering
\includegraphics[width=11.25cm,height=9.0cm]{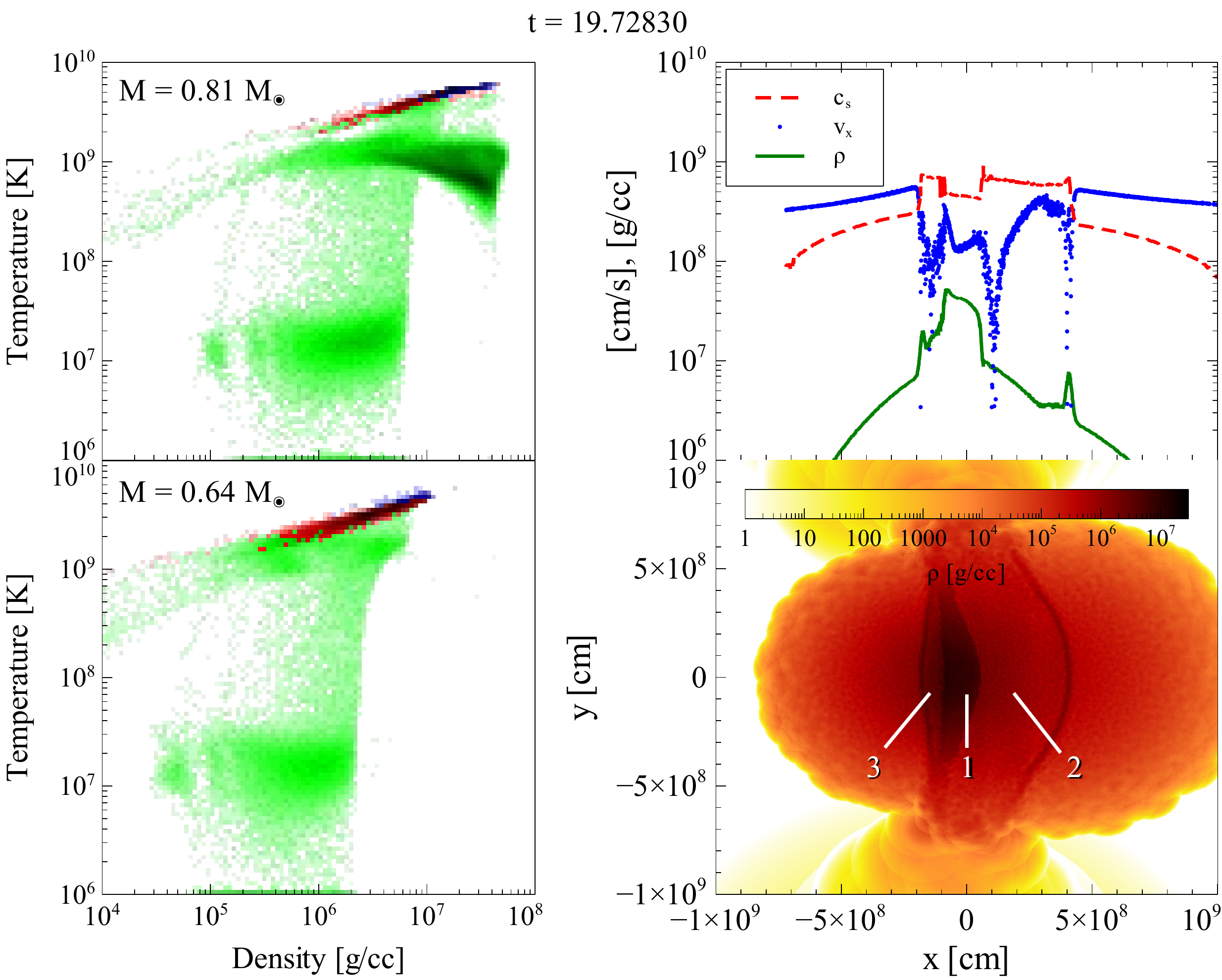}
\caption{Same format as Figure \ref{fig:1450} for mass pair 2, 0.64\msol + 0.81\msol, and $b=0$, at a later time in the simulation. Material shocked by the collision is labeled as region 1. Material behind the first detonation shock is labeled as region 2, and material behind the second detonation shock is labeled region 3.}
\label{fig:2750}
\end{figure*}

However, region 1 does not maintain its lenticular shape as the two stars are moving at different speeds relative to this shocked region. This allows the detonation shock to travel through this region at $\approx$ Mach 1, eventually reaching fresh carbon outside of region 1. This fresh carbon ignites explosively, creating a second detonation (region 3 in Figure \ref{fig:2750}), which sends leading shocks back through region 1 and into region 2, shocking it a second time and eventually catching up with the first detonation shock.

Most of the \nickel[56] in this scenario is produced in the more massive star, as Table \ref{table:pair2} demonstrates. Since only low-density portions of the 0.64\msol star had entered the shocked region before the detonation, most of its contribution to the total output is in Si-group elements.

\begin{table}[ht]
\caption{\rm{Isotope yields for the $b=0$ and $b=1$ cases of mass pairing 2, 0.64\msol + 0.81\msol.}}
\centering
\begin{tabular}{l | l | c c | c}
\hline\hline
$b$ & Isotope & 0.81 [\msol] & 0.64 [\msol] & Total [\msol]\\
\hline
\multirow{4}{*}{0} & \carbon[12] & 0.21 & 0.03 & 0.24\\
& \oxygen[16]	& 0.25 & 0.14 & 0.39\\
& \silicon[28]	& 0.12 & 0.27 & 0.39\\
& \nickel[56]	& 0.13 & 0.02 & 0.14\\
\hline
\multirow{4}{*}{1} & \carbon[12] & 0.02 & 0.03 & 0.05\\
& \oxygen[16]	& 0.07 & 0.14 & 0.21\\
& \silicon[28]	& 0.12 & 0.25 & 0.37\\
& \nickel[56]	& 0.49 & 0.04 & 0.53\\
\hline
\end{tabular}
\label{table:pair2}
\end{table}

\begin{figure}[ht]
\centering
\includegraphics[width=7cm,height=7cm]{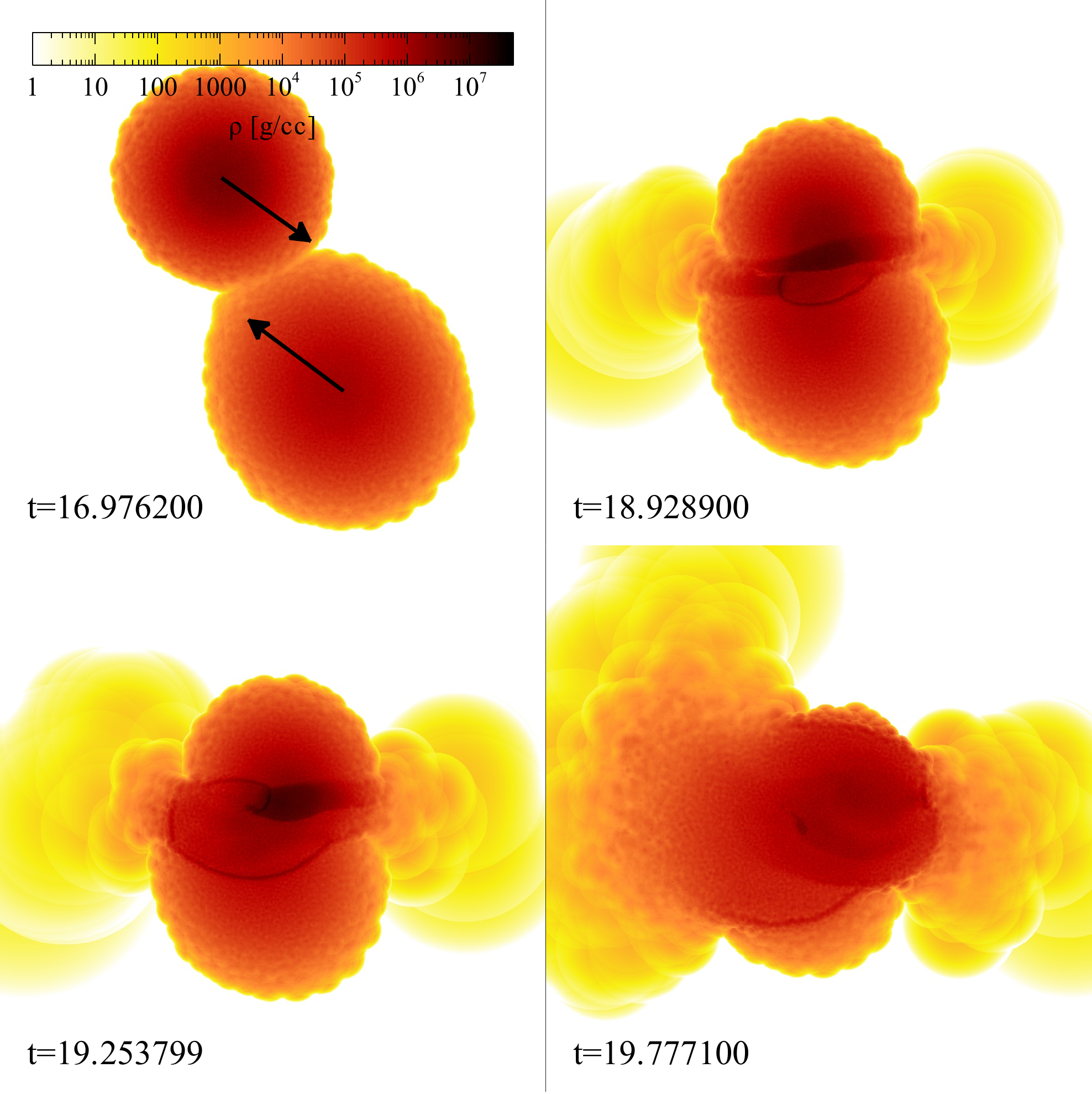}
\caption{A 2D slice of interpolated densities through the $x$-$y$ plane of the $b=1$ case of a 0.64\msol white dwarf colliding with a 0.81\msol white dwarf. Four snapshots at different times are shown. Arrows in the top-left panel indicate the directions of motion of each star.}
\label{fig:0p6_0p8_half}
\end{figure}

In the $b=1$ case, the pre-detonation, shocked region reaches much higher densities, and the oblique angle at which the white dwarf stars enter the shocked region allows more material to become strongly shocked by the detonation. The detonation shock also twists around the peculiar density contours inside the shocked region, shocking much of the material several times, as is seen in the bottom-right panel of Figure \ref{fig:0p6_0p8_half}. The 0.81\msol star experiences a near complete burn of all of its carbon and oxygen. However, as in the $b=0$ case, most of the 0.64\msol star remains unshocked at the time of the detonation breakout. As before, the 0.64\msol star contributes mostly Si-group elements to the total output, as shown in Table \ref{table:pair2}.

\begin{figure}[ht]
\centering
\includegraphics[width=5.25cm,height=6cm]{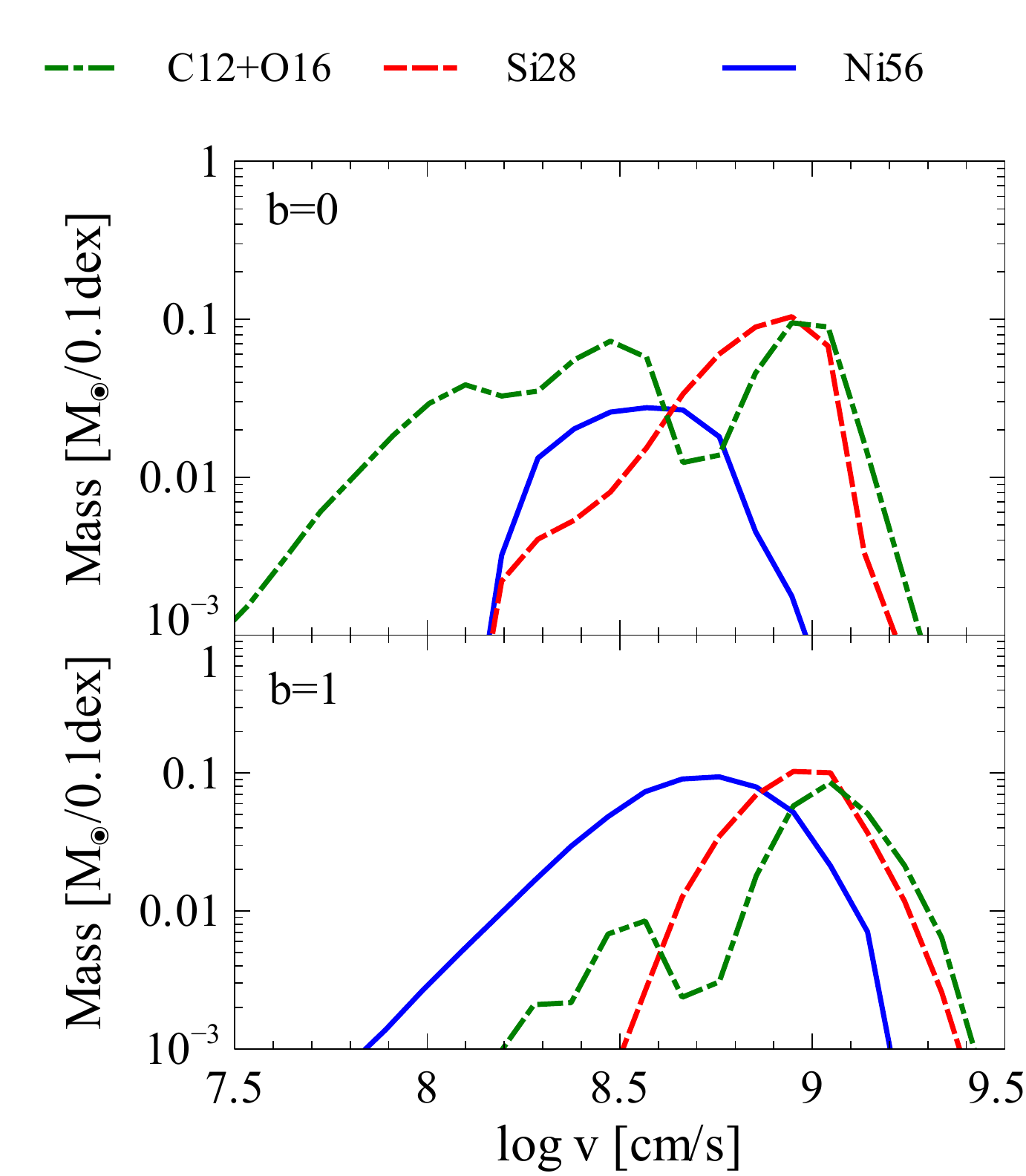}
\caption{Masses of several isotopes at logarithmically spaced velocity bins for the $b=0$ and $b=1$ cases of mass pairing 2, 0.64\msol + 0.81\msol.}
\label{fig:vel2}
\end{figure}

The velocity profiles for the $b=0$ and $b=1$ cases of mass pair 2, shown in Figure \ref{fig:vel2}, reinforce the observation that \nickel[56] is created in a confined region in the $b=0$ case, mainly in the densest portions of the shocked material from the 0.81\msol star. Carbon and oxygen, together, dominate the total output by mass, while in the $b=1$ case, \nickel[56] is the dominant isotope, followed by \silicon[28]. 

\subsection{Mass Pair 3 - 0.64\msol + 1.06\msol}

\begin{figure*}
\centering
\includegraphics[width=11.25cm,height=9.0cm]{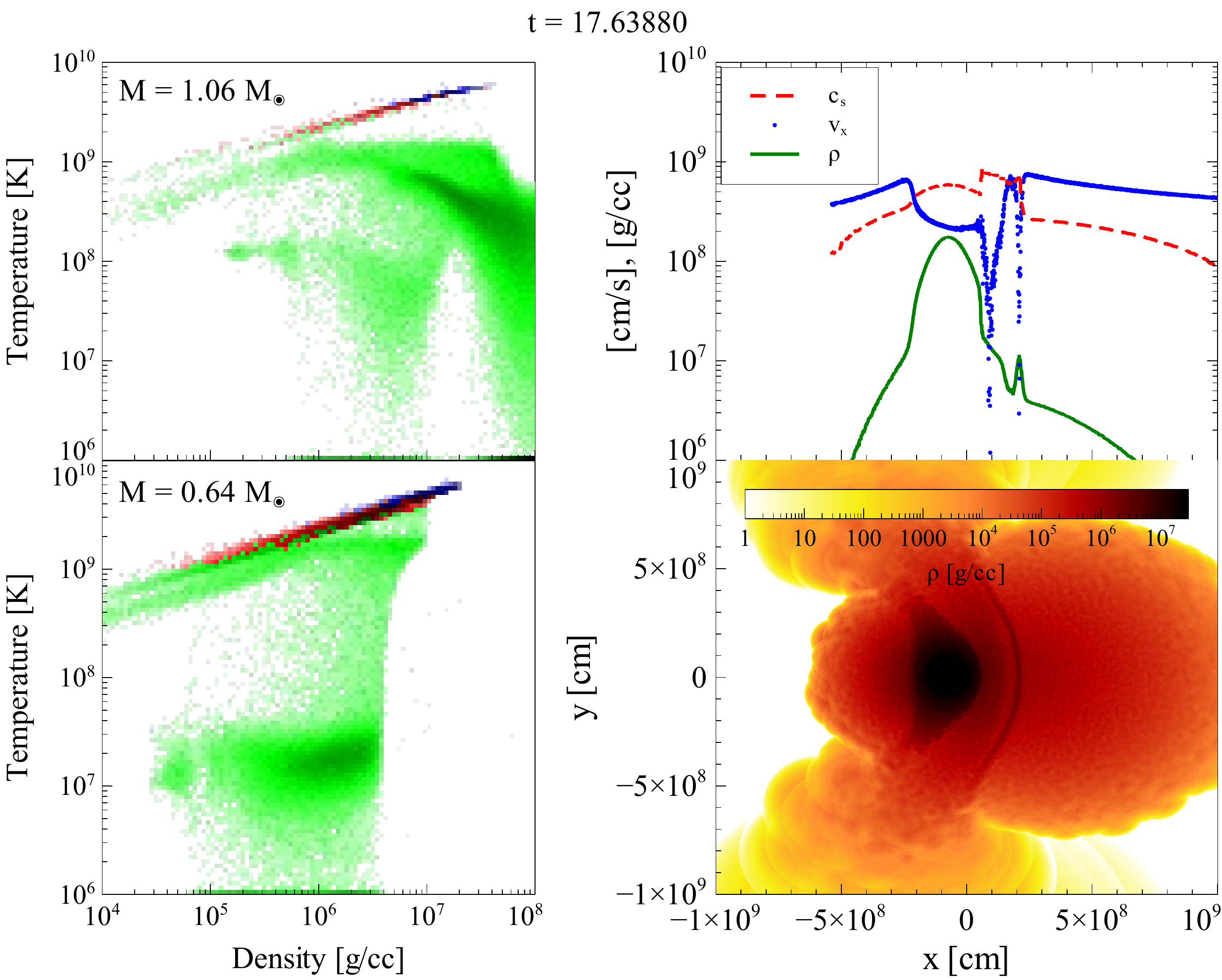}
\caption{Same format as Figure \ref{fig:1450} for mass pair 3, 0.64\msol + 1.06\msol, and $b=0$.}
\label{fig:4650}
\end{figure*}

As with mass pair 2, the $b=0$ case of mass pair 3 experiences a detonation of material in the 0.64\msol star very quickly after the stars first collide. However, owing to the greater potential well into which the 0.64\msol star is falling, the sound speed of the material shocked by the detonation is less than the infall velocity as shown in the top-left panel of Figure \ref{fig:4650}. 

After $\approx 0.7$s, as the core of the 1.06\msol star enters the shocked region, a second detonation lights on the left edge of the shocked region. This powers a shock that travels through both stars, catching up with the shock from the first detonation in the 0.64\msol star. The material in the 0.64\msol star burns mostly to \silicon[28], while what burns in the 1.06\msol star burns almost entirely to \nickel[56], due to its higher density. The contributions from each star to the total elemental abundances are given in Table \ref{table:pair3}.

\begin{table}[ht]
\caption{\rm{Isotope yields for the $b=0$ case of mass pairing 3, 0.64\msol + 1.06\msol.}}
\centering
\begin{tabular}{l | c c | c}
\hline\hline
Isotope & 1.06 [\msol] & 0.64 [\msol] & Total [\msol]\\
\hline
$^{12}$C	& 0.39 & 0.02 & 0.41\\
$^{16}$O	& 0.40 & 0.15 & 0.55\\
$^{28}$Si	& 0.05 & 0.24 & 0.29\\
\nickel[56]	& 0.19 & 0.07 & 0.26\\
\hline
\end{tabular}
\label{table:pair3}
\end{table}

In simulations of mass pair 3 that introduced a non-zero impact parameter, the 1.06\msol star was simply too compact to be significantly disrupted by a collision with a 0.64\msol star. In both the $b=1$ and $b=2$ cases, most of the 1.06\msol star survived the collision, while completely disrupting the 0.64\msol star. Details of those simulations are given in \S4.


\subsection{Mass Pair 4 - 0.81$\msol\times2$}

\begin{figure*}
\centering
\includegraphics[width=11.25cm,height=9.0cm]{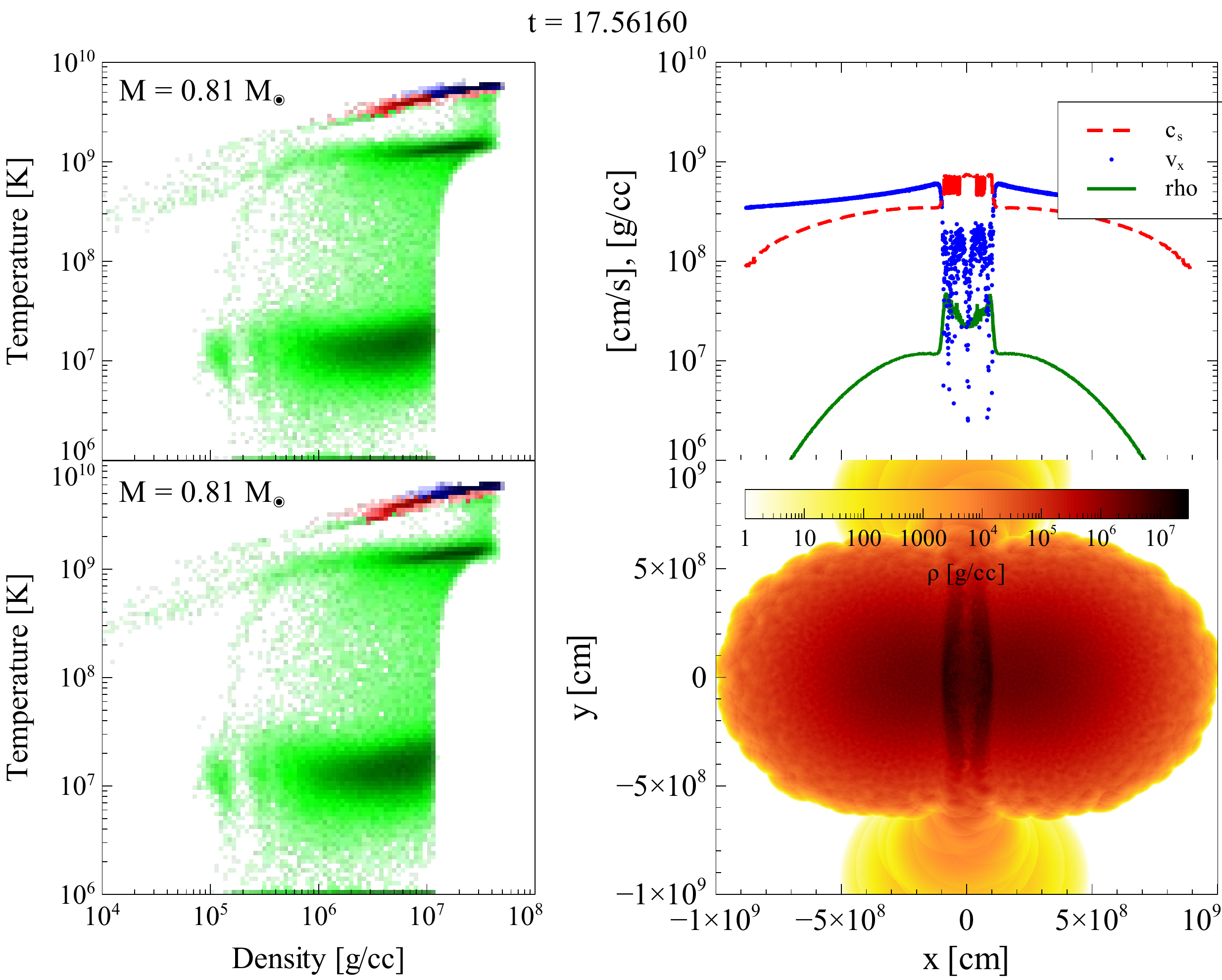}
\caption{Same format as Figure \ref{fig:1450} for mass pair 4, 0.81\msol$\times2$, and $b=0$.}
\label{fig:2300}
\end{figure*}

The symmetrical mass pair, 0.81\msol$\times2$ is quite unlike the 0.64\msol$\times2$ mass pairing discussed above. For the 0.81\msol$\times2$ mass pair with $b=0$, several detonations occur in the $y$-$z$ plane simultaneously and almost immediately after impact, owing to the higher temperatures reached in the shocked region from the higher infall speeds. As Figure \ref{fig:2300} shows, these detonations superimpose and produce copious amounts of \nickel[56] as they travel through the much denser material present inside the 0.81\msol stars. This denser material allows for a significantly greater conversion of carbon and oxygen to \nickel[56]. Therefore, with only a 26\% increase in total mass of the system over the 0.64\msol$\times2$ scenario, there is a 64\% increase in \nickel[56] production to 0.84\msol. 

Having denser and more compact stars also reduces the sensitivity of the \nickel[56] yield to impact parameter. Indeed, with two 0.81\msol stars, both the $b=1$ and $b=2$ simulations resulted in detonations and significant \nickel[56] production, 0.84\msol and 0.65\msol, respectively. Differences in the \nickel[56] yield for the two non-zero impact parameter simulations stem mainly from the amount of material that burns to $^{28}$Si before the detonations occur, with the $b=2$ scenario featuring much more material burning at lower temperatures to silicon before the detonation. The high activation energy of $^{28}$Si prevents much of that material from being converted to \nickel[56].


\subsection{Mass Pair 5 - 0.81\msol + 1.06\msol}

The 0.81\msol + 1.06\msol mass pair follows a very similar pattern to that of mass pair 2 (0.64\msol + 0.81\msol). Intermediate impact parameters allow more material to enter the shocked region before detonation, and so there is a rise in \nickel[56] production in the $b=1$ case over $b=0$. However, because both stars involved in the collision are denser than their counterparts in mass pair 2, much more \nickel[56] is produced overall. Contributions to the total yield in the $b=0$ and $b=1$ simulations are given in Table \ref{table:pair5} below.

\begin{table}[ht]
\caption{\rm{Isotope yields for the $b=0$ and $b=1$ cases of mass pairing 5, 0.81\msol + 1.06\msol.}}
\centering
\begin{tabular}{l | l | c c | c}
\hline\hline
$b$ & Isotope & 1.06 [\msol] & 0.81 [\msol] & Total [\msol]\\
\hline
\multirow{4}{*}{0} & $^{12}$C & 0.17 & 0.01 & 0.18\\
& $^{16}$O	& 0.19 & 0.09 & 0.28\\
& $^{28}$Si	& 0.06 & 0.22 & 0.28\\
& \nickel[56]	& 0.58 & 0.32 & 0.90\\
\hline
\multirow{4}{*}{1} & $^{12}$C & 0.05 & 0.02 & 0.07\\
& $^{16}$O	& 0.07 & 0.09 & 0.16\\
& $^{28}$Si	& 0.06 & 0.22 & 0.28\\
& \nickel[56]	& 0.82 & 0.31 & 1.13\\
\hline
\end{tabular}
\label{table:pair5}
\end{table}

\subsection{Mass Pairs 6 \& 7 - 0.96$\msol\times2$ \& 1.06\msol$\times2$}

The 0.96\msol$\times2$ and 1.06\msol$\times2$ simulations were essentially similar to the 0.81\msol$\times2$ simulations with the exception that the greater the mass of the constituent stars, the less sensitive the \nickel[56] yield was to impact parameter. Indeed, both mass pairs 6 and 7 produced almost the same yield in all three tested collision scenarios. 

What distinguishes the 1.06\msol$\times2$ mass pair from all the others attempted is that the \nickel[56] yield is super-Chandrasekhar in all cases. Were such explosions observed, there would be no doubt that a double-degenerate progenitor scenario of some kind was responsible. The resulting 1.71\msol of \nickel[56] from the 1.06\msol$\times2$ simulations appear strikingly similar to the 1.7\msol of \nickel[56] derived from the observations of Scalzo \etal (2010). 

\subsection{Mass Pair 8 - 0.50$\msol\times2$}

Finally, we studied symmetric collisions of low-mass, 0.50\msol white dwarfs. Table \ref{table:runs} demonstrates that the $b=0$ collision scenario for this mass pairing produced less than $0.01\msol$ of \nickel[56] despite having resulted in a detonation. In this case, the energy generated from even mild carbon-burning was sufficient to unbind the stars. As in the 0.64\msol$\times2$, $b=0$ scenario, the lower velocities with which the stars collide results in a late detonation. The shocked region slowly heats up until carbon-burning at its edges ignites a detonation. 

\begin{figure}[ht]5.25
\centering
\includegraphics[width=5.25cm,height=3.75cm]{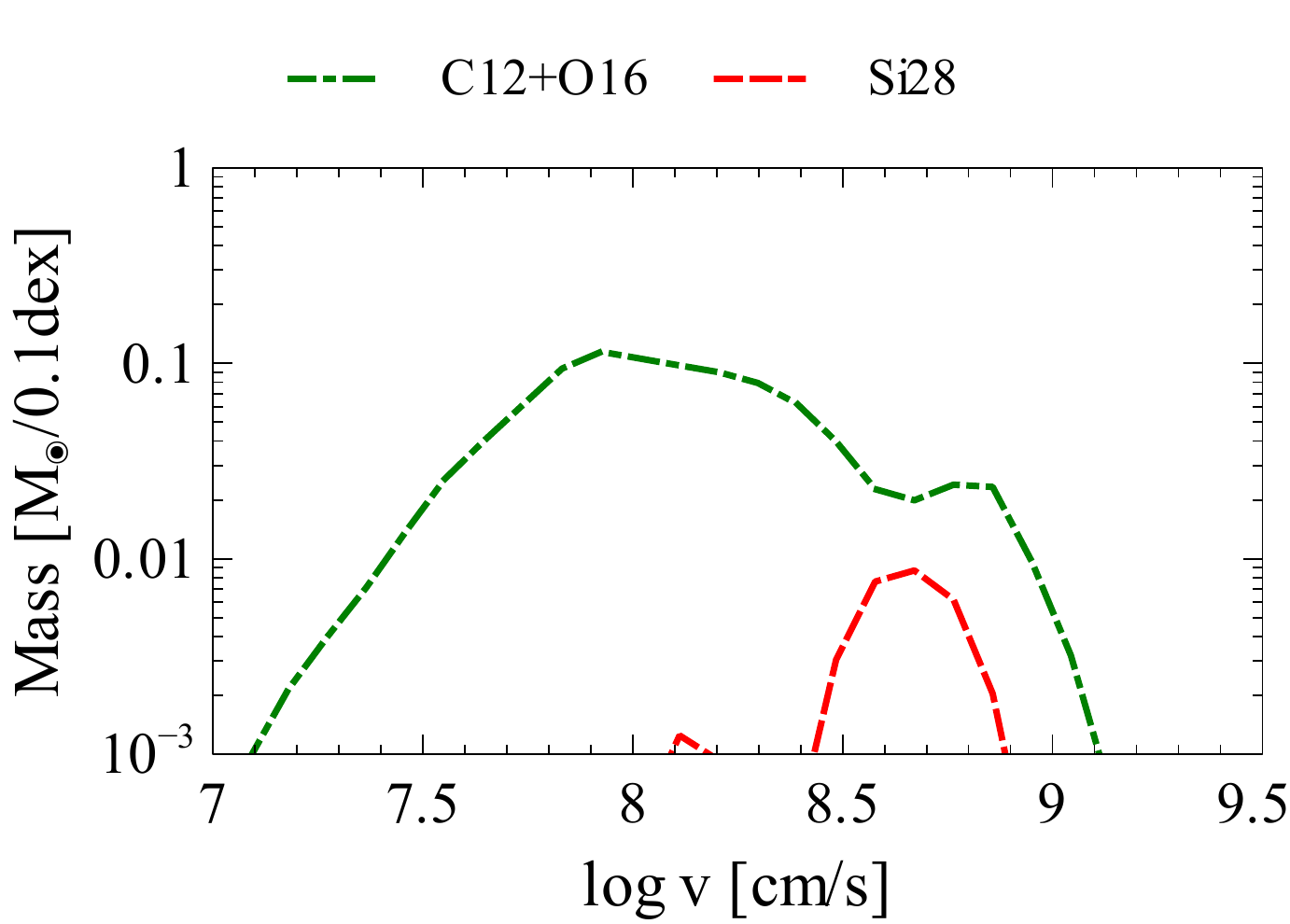}
\caption{Masses of several isotopes at logarithmically spaced velocity bins for the $b=0$ case of mass pairing 8, 0.50\msol$\times2$.}
\label{fig:vel8}
\end{figure}

It is clear from the velocity profile of the most abundant isotopes from this collision, given in Figure \ref{fig:vel8}, that carbon and oxygen remain mostly unburned in this scenario. This seems to suggest that collisions of low-mass white dwarfs ($M\apprle0.6\msol$) of the CO variety would not produce observable transients. Other simulations introducing impact parameters with this mass pair were not attempted with carbon-oxygen white dwarfs as the $b=0$ simulation yielded essentially a non-result. However, further investigation involving Helium white dwarfs is warranted. 

\section{Results \& Analysis II - Remnants}

\begin{figure*}
\centering
\includegraphics[width=16cm,height=10.66cm]{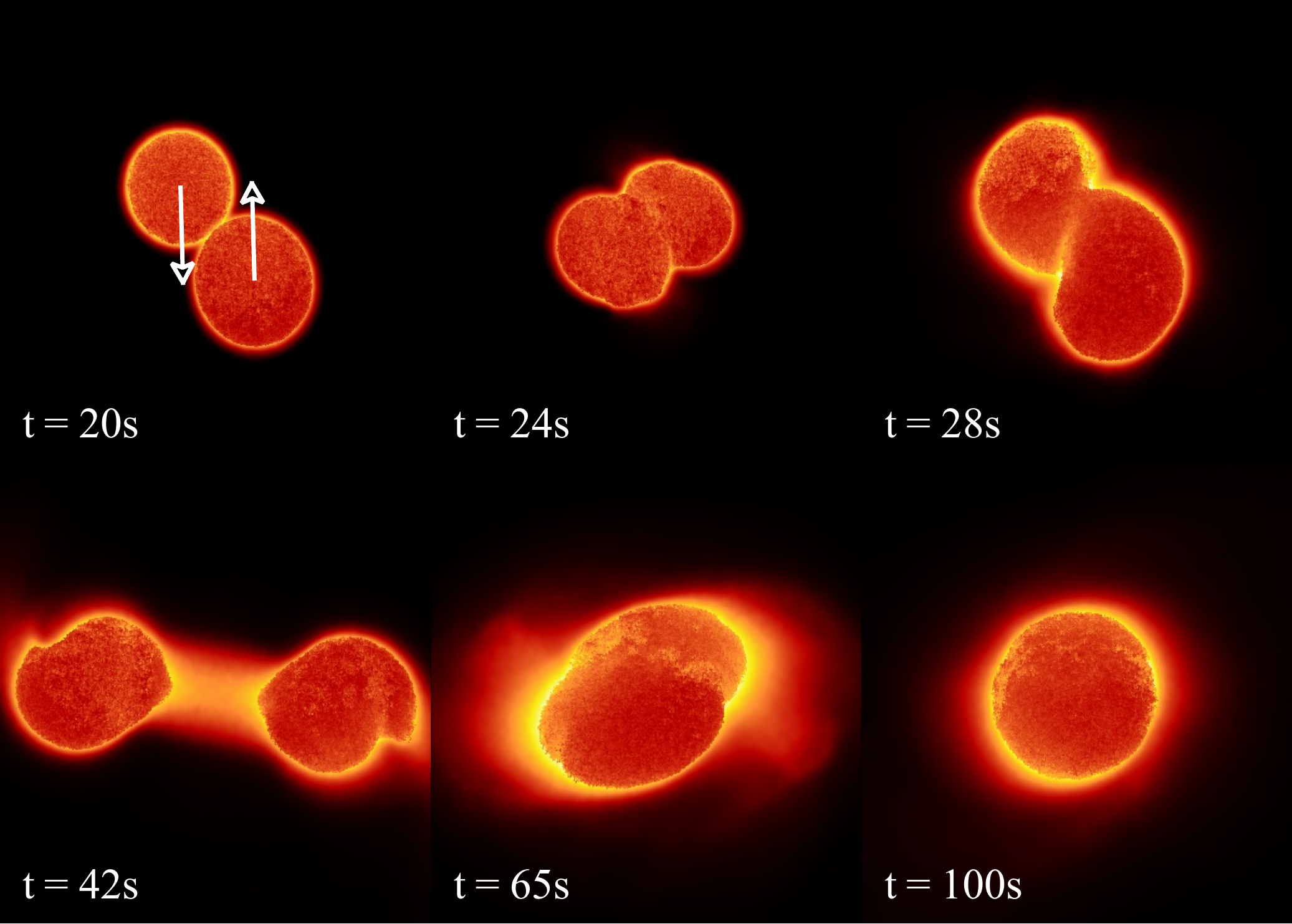}
\caption{Snapshots of density isosurfaces at six different times for the $b=2$ simulation of mass pair 1, 0.64\msol$\times2$. After first colliding, the stars separate before coalescing into a single object.}
\label{fig:0p6x2_full}
\end{figure*}

As seen in Raskin \etal (2009), the $b=2$ case of mass pair 1 (0.64\msol $\times 2$) did not feature a detonation and instead formed a hot remnant of thermally-supported carbon and oxygen with some carbon-burning products. Figure \ref{fig:0p6x2_full} illustrates the dynamics of this collision, starting with a glancing case that leads to the constituent stars spinning off from each other before coalescing into a single hot object.

The compact remnant core after 100s featured a nearly constant density of $\rho \sim 10^6$ g cm$^{-3}$. It was surrounded by a thick, Keplerian disk $\approx 2.0\times10^{10}$ cm in radius with a scale radius $r_0 \approx 2.3\times10^9$ cm. The compact object at the center of the disk is not strictly a white dwarf since much of its pressure support is thermal ($T\approx 5\times 10^8$ K). Indeed, since degeneracy pressure support necessitates that more massive white dwarfs are smaller than less massive ones, this object, at $\approx 0.8$\msol is far too large to be wholly degenerate ($r_{rem}\approx2.5\times10^9$ cm); larger than the 0.64\msol white dwarfs that entered into the collision ($r_{0.64}=6.98\times10^8$ cm).

Carbon ignition nominally takes place at approximately 7-8$\times$10$^8$ K (\eg Gasques \etal 2007), but recent phenomenological models (\eg Jiang \etal 2007) have suggested a strongly reduced, low-energy astrophysical S-factors for carbon fusion reactions that potentially reduce carbon ignition temperatures to $\approx$ 3$\times$10$^8$ K, especially at densities of 10$^9$ g cm$^{-3}$. A lower carbon burning threshold would be of interest to future studies of collision remnants.

Since the system started in a bound state, ($T\lesssim-V$, where $T$ in this case is total kinetic energy and $V$ is total gravitational potential energy) and since any energy gained from nuclear processes is negligible, most of the material cannot escape the system and the disk remains bound to the compact core. It will eventually cool and collapse onto the surface of the compact object. However, the hot core may accelerate parts of the disk to escape velocity via radiative processes, and so the calculation of the final mass of the resultant white dwarf is beyond the scope of this paper. Suffice it to say, the final mass will not exceed the Chandrasekhar limit as only 1.28\msol of material is available. 

For the $b=2$ case of mass pair 2, 0.64\msol + 0.81\msol, the compact remnant was slightly less massive at $\approx0.75$\msol. However, since the total mass of the system is super-Chandrasekhar, the final remnant mass may result in a super-Chandrasekhar white dwarf. Again, this final mass will depend greatly on radiative processes, and the likelihood of producing a SNIa will hinge on the accretion rate of the disk onto the core. 

\begin{figure}[ht]
\centering
\includegraphics[width=7cm,height=6cm]{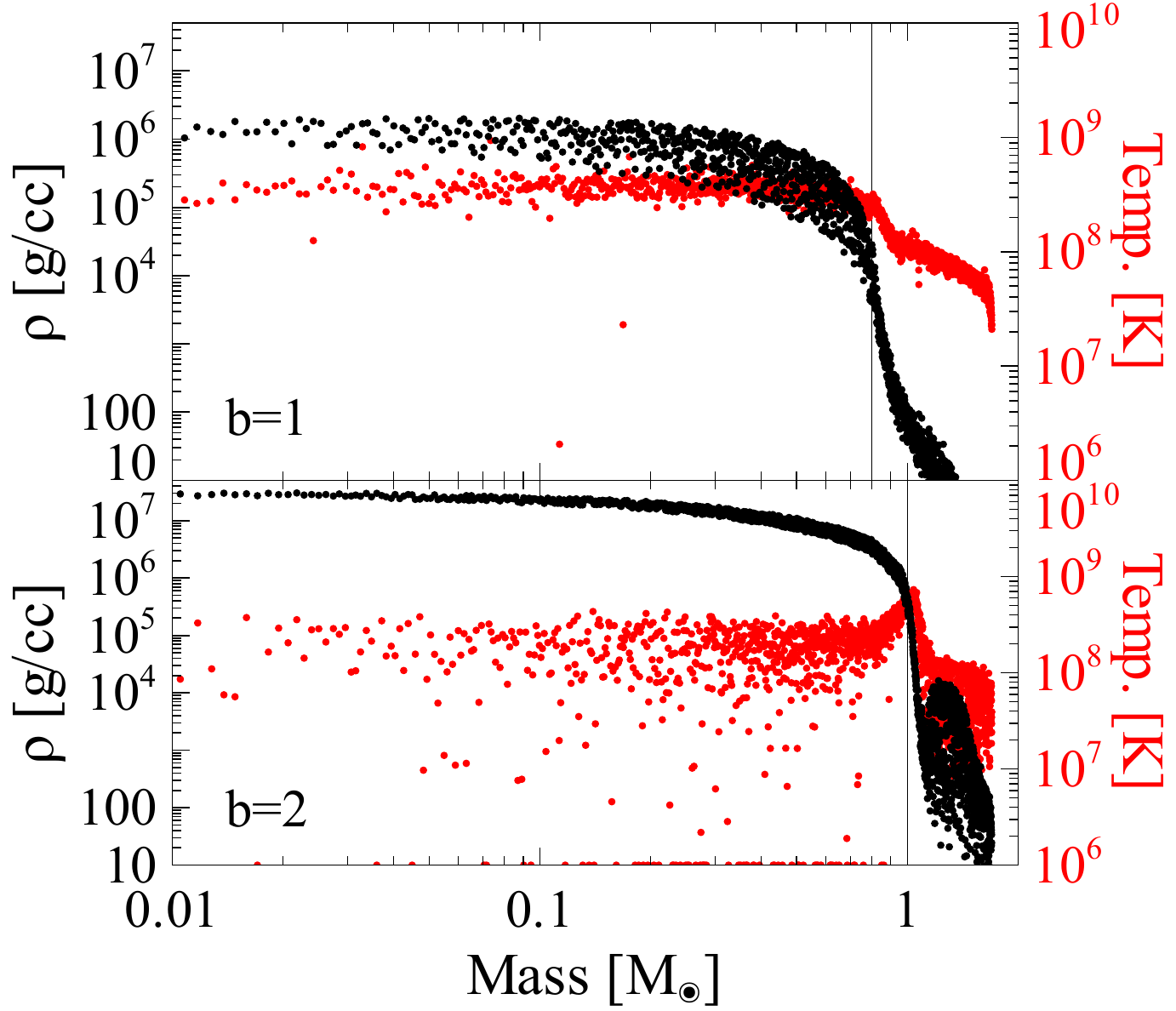}
\caption{Temperature and density profiles of the $b=1$ and $b=2$ simulation remnants of mass pair 3, $0.64\msol + 1.06\msol$.}
\label{fig:mp3rem}
\end{figure}

The simulations of mass pair 3, 0.64\msol + 1.06\msol, resulted in remnants in both the $b=1$ and $b=2$ cases as the 1.06\msol white dwarf was too compact for the star to be much affected by a grazing collision with a 0.64\msol white dwarf. In the $b=1$ case, some of the atmosphere of the 1.06\msol star was stripped away to join the material from the disrupted 0.64\msol star in the disk, while in the $b=2$ case, the 1.06\msol star was nearly unaffected by the collision. Figure \ref{fig:mp3rem} illustrates that the remnant core of the $b=2$ simulation masses $\approx1.0\msol$, and the central densities are essentially unchanged from the 1.06\msol progenitor. 

Some features worth noting in the $b=2$ profiles in Figure \ref{fig:mp3rem} are the indications of a cold core ($T_{\rm core}\approx 2\times 10^8$ K) surrounded by a hot envelope ($T_{\rm env.}\approx 9\times 10^8$ K), and the presence of a strong overdensity in a part of the disk, which causes large spreads in density and temperature for the disk material. This suggests that while the 0.64\msol star is no longer a gravitationally bound object, it has nevertheless not been completely disrupted.

\begin{figure}[ht]
\centering
\includegraphics[width=7cm,height=4cm]{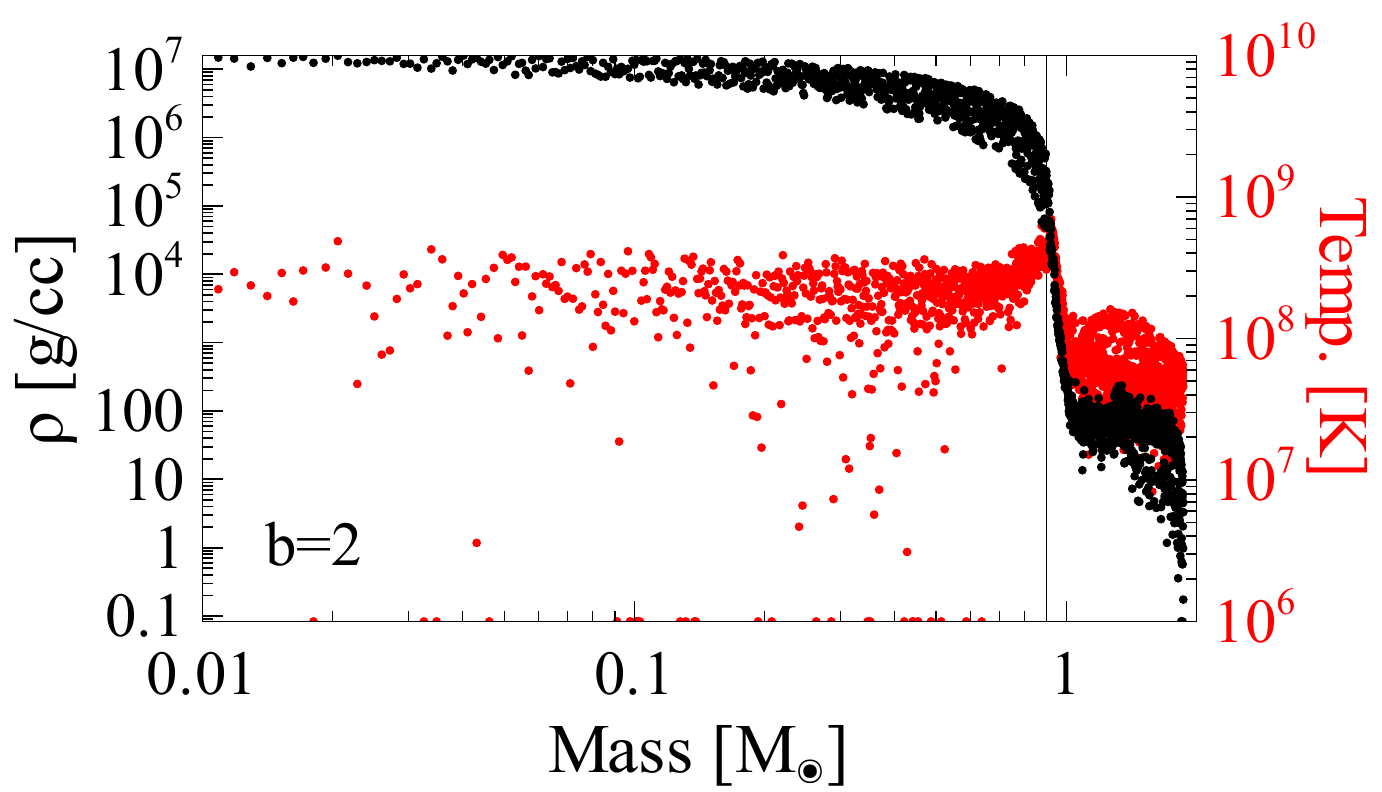}
\caption{Temperature and density profiles of the $b=2$ simulation remnant of mass pair 5, $0.81\msol + 1.06\msol$.}
\label{fig:mp5rem}
\end{figure}

The radial profiles of the $b=2$ simulation of the 0.81\msol + 1.06\msol mass pair, shown in Figure \ref{fig:mp5rem}, indicate a very similar structure, with a high-temperature envelope surrounding a cold core and an overdensity in a part of the disk. These inhomogeneities in the disk would most likely vanish after many crossing-times, however, it is highly unlikely that radiative processes or collisional excitations within the disk could remove as much as 0.4\msol. This suggests that after a Kelvin-Helmholtz timescale, the remnant from the $b=2$ collision of mass pair 5 will almost certainly be super-Chandrasekhar. 

Some of the properties of these remnants are very similar to the simulated remnant explored in Yoon \etal (2007), which was produced by the merger of 0.9\msol and 0.6\msol white dwarfs. In that simulation, the remnant also featured a cold core ($T_{\rm core}\approx 1\times 10^8$ K) surrounded by a hot envelope ($T_{\rm env.}\approx 6\times 10^8$ K), embedded inside a thick, Keplerian disk. Using a 1D stellar evolution code, they found that such systems can indeed evolve on timescales $\sim10^5$yr toward becoming SNeIa.

\section{Discussion}

White dwarf collisions are not typically regarded as SNeIa progenitors, and therefore, they have been relatively unexplored theoretically.  Here we have conducted a comprehensive suite of simulations of such collisions examining the dependence of their \nickel[56] yield on total mass, mass ratio, and impact parameter.  Our results suggest that white dwarf collisions are a viable avenue for producing SNeIa with brightnesses that range from sub-luminous to super-luminous.

In fact, in more than 75\% of our simulations, collisions resulted in detonations, and in all but the least massive combination of stars, significant amounts of \nickel[56] were produced. We found that even mass pairs that are below the Chandraskehar limit featured explosive nuclear burning, with the 0.64\msol$\times2$ mass pair producing \nickel[56] in quantities comparable to standard SNeIa. Moreover, the most massive combinations of stars produced super-luminous quantities of \nickel[56], regardless of the impact parameter, greatly increasing their likelihood of detection. The \nickel[56] yields from these collisions are consistent with those of observed SNeIa with super-Chandrasekhar mass progenitors.

Asymmetric mass pairs generally produced less \nickel[56] in head-on collisions than symmetric pairs. At middling impact parameters, much more \nickel[56] was produced, however, at high impact parameters, there was little or none. This is due primarily to the delicate balance which must be struck between the dynamics of the impact and the binding energy of the less massive star in order to establish a stalled shock region that can lead to a detonation. At high impact parameters, the less massive star is typically unbound by the collision before much or any \nickel[56] is produced.

For combinations of masses and impact parameters that did not detonate, the end result always featured a compact, semi-degenerate object surrounded by a bound, thick disk of carbon and oxygen. Many of these systems were super-Chandrasekhar, and over Kelvin-Helmholtz time scales, these, too, are candidate progenitors for producing SNeIa.

Our results have shown that \nickel[56] production in white dwarf collisions is a non-linear process that depends on several factors, including infall velocities and tidal distortion effects.  Foremost among parameters to be explored in future studies is the composition of the constituent white dwarfs.  Helium has a much lower activation energy than carbon or oxygen, and combinations of stars that include helium white dwarfs would almost certainly produce interesting and different results.  Other avenues to be explored include the impact of more detailed modeling of the isotopic profiles in the progenitor stars, and the possibility of sparse hydrogen atmospheres.  The results of these studies will shed further light on the contribution of double-degenerate collisions to the observed population of SNeIa.

\section*{Acknowledgments}

This work was supported by the National Science Foundation under grant AST 08-06720, by the National Aeronautics and Space Administration under NESSF grant PVS0401, and by a grant from the Arizona State University chapter of the GPSA. All simulations were conducted at the Ira A. Fulton High Performance Computing Center at Arizona State University. We thank James Rhoads and Sumner Starrfield for insightful discussions and our anonymous referee for useful suggestions and feedback.

\end{document}